\documentclass[lettersize,journal]{IEEEtran}
\usepackage{amsmath,amsfonts}
\usepackage{algorithmic}
\usepackage{algorithm}
\usepackage{array}
\usepackage[caption=false,font=normalsize,labelfont=sf,textfont=sf]{subfig}
\usepackage{textcomp}
\usepackage{stfloats}
\usepackage{url}
\usepackage{verbatim}
\usepackage{graphicx}

\usepackage{bbding}
\usepackage{cite}
\makeatletter
\def\UrlAlphabet{%
      \do\a\do\b\do\c\do\d\do\e\do\f\do\g\do\h\do\i\do\j%
      \do\k\do\l\do\m\do\n\do\o\do\p\do\q\do\r\do\s\do\t%
      \do\u\do\v\do\w\do\x\do\y\do\z\do\A\do\B\do\C\do\D%
      \do\E\do\F\do\G\do\H\do\I\do\J\do\K\do\L\do\M\do\N%
      \do\O\do\P\do\Q\do\R\do\S\do\T\do\U\do\V\do\W\do\X%
      \do\Y\do\Z}
\def\UrlDigits{\do\1\do\2\do\3\do\4\do\5\do\6\do\7\do\8\do\9\do\0}
\g@addto@macro{\UrlBreaks}{\UrlOrds}
\g@addto@macro{\UrlBreaks}{\UrlAlphabet}
\g@addto@macro{\UrlBreaks}{\UrlDigits}
\makeatother

\begin{document}

\title{FlatProxy: A DPU-centric Service Mesh Architecture for Hyperscale Cloud-native Application}

\author{Ming Li, Wenyan Lu, Hanyue Lin, Jingya Wu, Yu Zhang, Guihai Yan
        % <-this % stops a space
\thanks{The authors are with State Key Lab of Computer Architecture, Institute of Computing Technology, Chinese Academy of Sciences, Beijing, China, 100190. To whom correspondence should be addressed. E-mail: liming950814@gmail.com, \{luwenyan, yan\}@ict.ac.cn}% <-this % stops a space
\thanks{Manuscript received April 19, 2023; revised August 16, 2023.}}

\markboth{Journal of Transactions on Computers,~Vol.~14, No.~8, August~2023}%
{Ming Li, Wenyan Lu \MakeLowercase{\textit{et al.}}: FlatProxy: A DPU-centric Service Mesh Architecture for Hyperscale Cloud-native Application}

\IEEEpubid{0000--0000/00\$00.00~\copyright~2023 IEEE}

\maketitle

\begin{abstract}
Service mesh is a fundamental technology for building cloud-native applications, which ensures the stable running of a large number of services by an intermediate layer that governs communication between services. However, service mesh is not well suited for high-performance scenarios. The root cause is that the current service mesh is not suitable for the evolution of cloud-native applications. On the one hand, the service mesh built on CPU cannot listen to communication bypassing the CPU. On the other hand, service mesh includes many I/O-intensive and computationally-intensive tasks that can overload CPU cores as traffic grows beyond CPU performance.

Therefore, we propose a data-centric service mesh that migrates the proxy of the service mesh to the entrance of the network. Moreover, we also design the DPU-centric FlatProxy, a data-centric service mesh based on DPU. There are three advantages to the DPU-centric service mesh. Firstly, it takes over all traffic flow in and out of the node, which expands the sense scale of the service mesh from container to node. Secondly, it improves communication performance and reduces host resource usage by offloading some functions and optimizing communication. Thirdly, it minimizes performance and security issues through the physical isolation of business services and cloud infrastructure.

 Compared with Envoy, the current mainstream service mesh implementation, FlatProxy reduces latency by 90\% and improves throughput by 4x in Gbps and 8x in qps, and it only occupies a small amount of CPU resources.
\end{abstract}


\section{Introduction}

\IEEEPARstart{I}{n} the cloud-native era, applications are deployed using the microservice architecture, which enhances the flexibility, scalability, and stability of the application  \cite{1.microservice_review0}. However, the running of a larger number of microservices poses challenges in service interactions \cite{10.soa_governance1}. To reduce the complexity of microservice cooperation, service mesh is proposed to ensure that plenty of services run steadily. Service mesh is an intermediate layer between the network and services, and this layer generally includes a set of proxies and each proxy deployed in the container is alongside with service containers in the same Pod (Pod is the minimum unit to share the storage and network resources between different containers \cite{77.pod}). Service mesh decouples governance functions, such as service discovery, traffic control, network observation, and security checking, from the service and implement these functions by proxy, which simplifies the development and deployment of cloud-native services \cite{25.istio}.

\begin{figure}[t]
\centering     %%% not \center
\subfloat[The CPU-centric proxy]{ 
    \includegraphics[width=115pt]{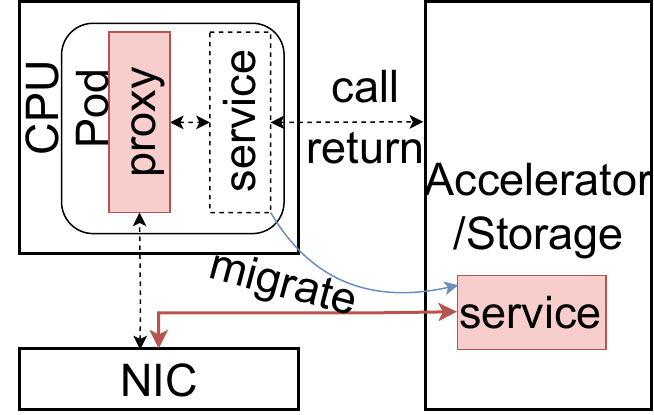}
    \label{fig:proxy_deployment:a}}
\subfloat[The DPU-centric proxy]{
    \includegraphics[width=115pt]{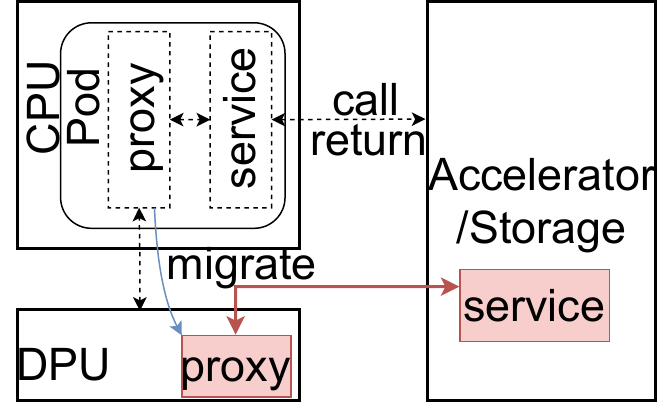}
    \label{fig:proxy_deployment:b}}

\caption{The deployment of proxy.}
\label{fig:proxy_deployment}
\end{figure}

However, with the emerging of data-centric cloud-native applications \cite{58.Lynx,62.lambda-NIC,63.NVMe-NIC,64.GPUDirect}, there are two serious problems for the current service mesh based on CPU. Firstly, the service mesh proxy is unable to listen to communication bypassing the CPU between the network and network-attached devices. As shown in Figure \ref{fig:proxy_deployment:a}, many services  are directly deployed on the accelerator or storage devices and they communicate with network bypassing CPU, but the traditional proxy deployed on the host listen to the communication by the Iptable mechanisms in the kernel, which makes proxy cannot sense this communication. Although some research has developed communication mechanisms between the CPU and NIC to monitor communication or offload the traffic manager to the device \cite{46.AccelTCP,66.smartDS}, diverse service governance will increase the system's complexity and governance costs with more services supported by different network-attached accelerators or storage.

\IEEEpubidadjcol
Secondly, the service mesh proxy includes many I/O intensive and computing intensive tasks that consume plenty of CPU resources and degrade communication performance \cite{12.serviceMesh_performance0, 67.serviceMesh_performance}. According to our experiments, deploying service mesh increases communication latency by more than 2x, decreases throughput by over 60\%, and consumes over 70\% of CPU resources. Previous works \cite{15.NSM-VPP, 26.cilium, 17.cloud-native-gateway} have optimized service mesh by building user-mode protocols, implementing kernel service mesh, and sharing network protocol stack or proxy, but these optimizations do not consider the current situation where CPU performance stalls while I/O bandwidth continues to increase. Some studies \cite{48.dagger, 46.AccelTCP, 51.panic} offload data processing functions, but they only accelerate network processing instead of service governance. Currently, there is no work on improving service mesh using hardware acceleration.

Therefore, we propose the data-centric service mesh, which migrates the proxy to the entrance of the network to implement service governance. As shown in Figure \ref{fig:proxy_deployment:b}, the data-centric service expands the scope of service governance from Pod to node and implements physical isolation between applications and cloud infrastructure, effectively eliminating host resource consumption used by the service mesh. However, expanding the government scope means more traffic needs to be processed, and physical isolation maybe cause performance loss due to the communication path being longer between service and proxy when the implementation of new service mesh is unreasonable \cite{67.serviceMesh_performance}. Some works \cite{68.DPFS} have tried to offload infrastructure software into embedded CPUs on the DPU (data processing unit), but we find that implementing the proxy on the embedded CPU is challenging to meet performance requirements because heavy data transmission can overload the weak embedded CPU.

To address the performance issue, we propose FlatProxy, a hardware-accelerated architecture of service mesh with a DPU. This design implements control and configuration functions by software and accelerates I/O-intensive and computation-intensive functions like routing, load-balancing, etc., by specific hardware. Additionally, in response to the challenges in traffic processing, function mapping, and communication isolation of the data plane, we propose a multi-pronged approach as following. Firstly, to balance data processing performance and flexibility, we adopt a hierarchy processing architecture which is suitable for different packet processing requirements. For example, L2, L3, and L4 are more regular and can be implemented by pipeline to improve processing throughput. In contrast, functions at L7 are implemented by multi-core architecture because data processing in this layer is diverse and requires more flexibility.
Secondly, to implement data processing and dynamic data flow mapping, we propose a hardware multi-thread model which is composed of a set of protocol processing modules (PPM). PPM implements functional expansion and dynamic combination through dynamic loading task instructions and flexible interconnections between network, host, accelerators, storage, etc.
Thirdly, to ensure the communication isolation, we further simplify communication paths between services and FlatProxy by using I/O virtualization and socket direct eliminating redundant data paths \cite{22.SRIOV, 32.socketdirect,33.socketdirect1}.

Our major contributions are summaried as following:
\begin{itemize}
    \item We firstly propose the data-centric service mesh and design a new architecture for the service mesh based on DPU.
    \item We adapt software/hardware co-design to reconstruct the service mesh and propose multiple methods, including hierarchy processing, hardware multi-thread models, and simplifying communication paths, to address performance issues induced by the concentration of proxy and the isolation with host.
    \item We implement the prototype on an FPGA, and experiments show that FlatProxy significantly improves service mesh performance. The latency of microservices communication is reduced by 90\% compared to current software implementation, throughput increases by 4x, and service mesh functions occupy little CPU resources.
\end{itemize}

\section{Background and Motivation}

In this section, we introduce the evolution of cloud-native applications and the characteristics of service mesh. In addition, we also analyze the limitations of service mesh and present our idea.

\subsection{The Data-centric Application}
In the data center, network-attached accelerators and storage have become important ways to improve system performance by bypassing CPU, which has inspired the development of data-centric applications.

\begin{figure}[t]
\centering     %%% not \center
\subfloat[The control-centric architecture]{ 
    \includegraphics[width=110pt]{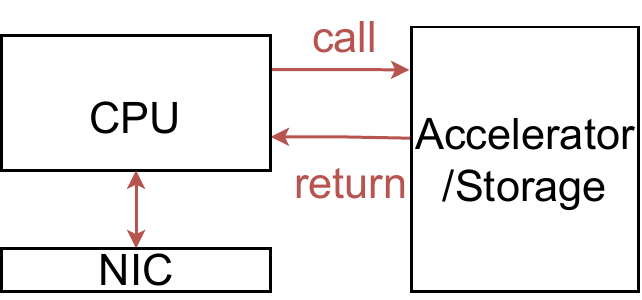}
    \label{fig:application_arch:a}}\quad
\subfloat[The data-centric architecture]{
    \includegraphics[width=110pt]{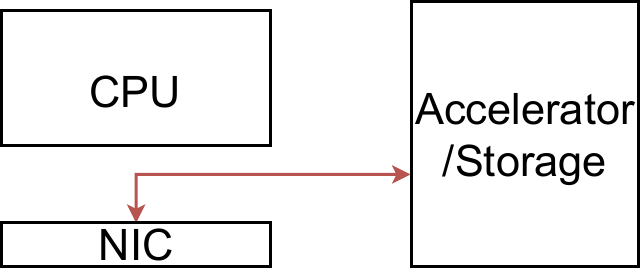}
    \label{fig:application_arch:b}}
\caption{The hardware-accelerated deployment of application.}
\label{fig:application_arch}
\end{figure}

As shown in figure \ref{fig:application_arch}, unlike control-centric applications that execute on the CPU and use the accelerator by call and return, data-centric cloud-native applications execute services on the accelerator and don't need to be controlled by the CPU \cite{45.accelNet, 63.NVMe-NIC, 58.Lynx, 70.serverless_on_heterogeneous}. For example, AccelNet and Nvlink show that the accelerator is directly attached to the NIC. Nvme-of shows that storage is directly attached to NIC. Lynx shows how to build data-centric software, and XPU-Shim shows the serverless architecture on heterogeneous architecture. The above works show a trend that services will be executed on different hardware according to their characteristics and independent of the CPU in the host.

\subsection{Service Mesh}
Cloud-native applications benefit from the microservices architecture, but ensuring stable operation for numerous services is a difficult problem. To address this issue, service mesh has been proposed, which decouples service governance functions from applications, forming a dedicated infrastructure layer to meet the requirements of service communication \cite{25.istio}.

The key functions of service mesh include three parts. The first is service discovery, which ensures that services can access others through the service mesh. The second is to ensure the robustness of service communication. Service mesh implements various traffic control functions such as load balancing, traffic limiting, circuit breaking, and more. The third is to maintain system observability to simplify fault diagnosis, traffic statistics, etc. In public cloud environments, security is also a necessary requirement.

\begin{figure}[t]
  \centering
  \includegraphics[width=200pt ]{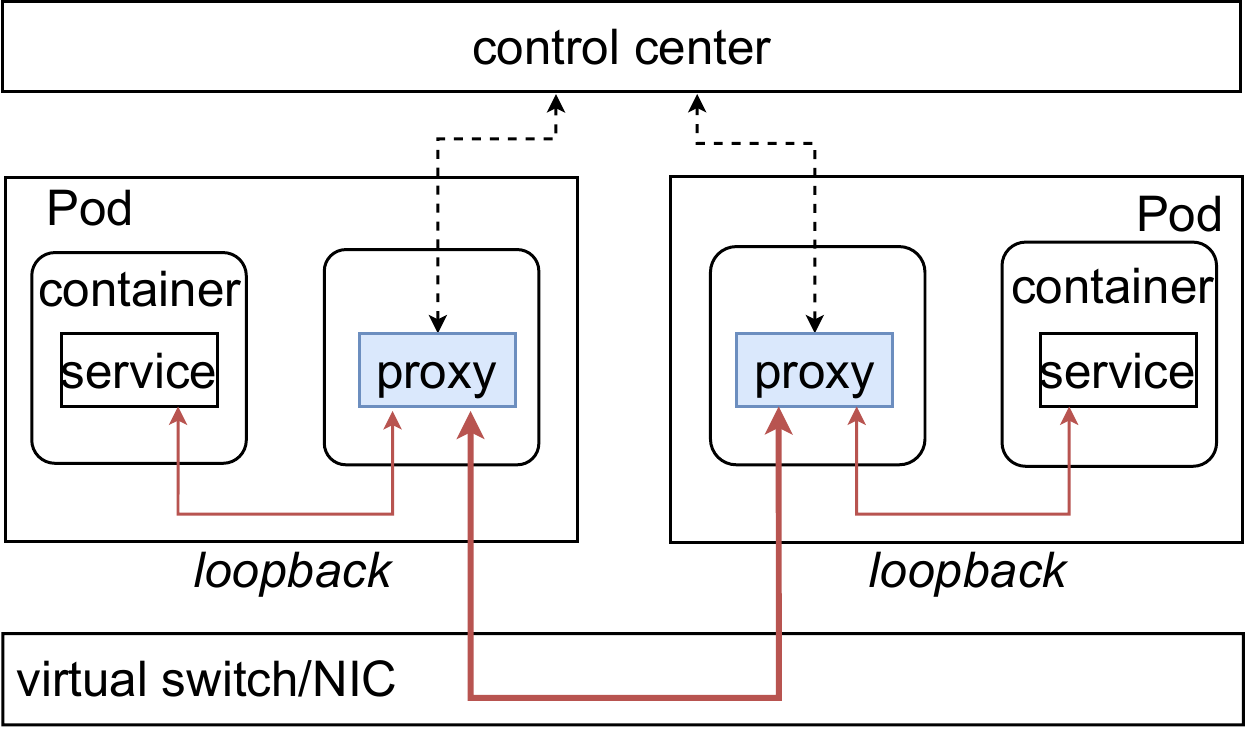}
  \caption{The architecture of service mesh}
  \label{fig:serviceMesh}
\end{figure} 

Currently, service mesh is implemented by a set of sidecar proxies deployed alongside services in the same POD (the minimum unit for sharing storage and network resources between different containers) and a control center to distribute configuration information \cite{24.Monolith, 25.istio}. As shown in figure \ref{fig:serviceMesh}, the sidecar proxy executes all functions of service governance and takes over all traffic in and out of the service. It communicates with services in the same Pod by loopback and with other services by network, which increases the original service communication path.

\subsection{Limitations of Service Mesh}
As stated above, we have identified three limitations with the implementation of service mesh on the Host.

\textbf{Inability to govern data-centric applications} When a service is deployed on a network-attached accelerator, the proxy will not sense the communication between the accelerator and the network because the sidecar proxy intercepts and captures traffic in and out of the service using Iptables \cite{25.istio}, a packet processing program in the kernel. This increases the complexity of service governance and limits the development of service mesh.

\begin{figure*}
\centering     %%% not \center
\subfloat[]{ 
        \includegraphics[width=110pt]{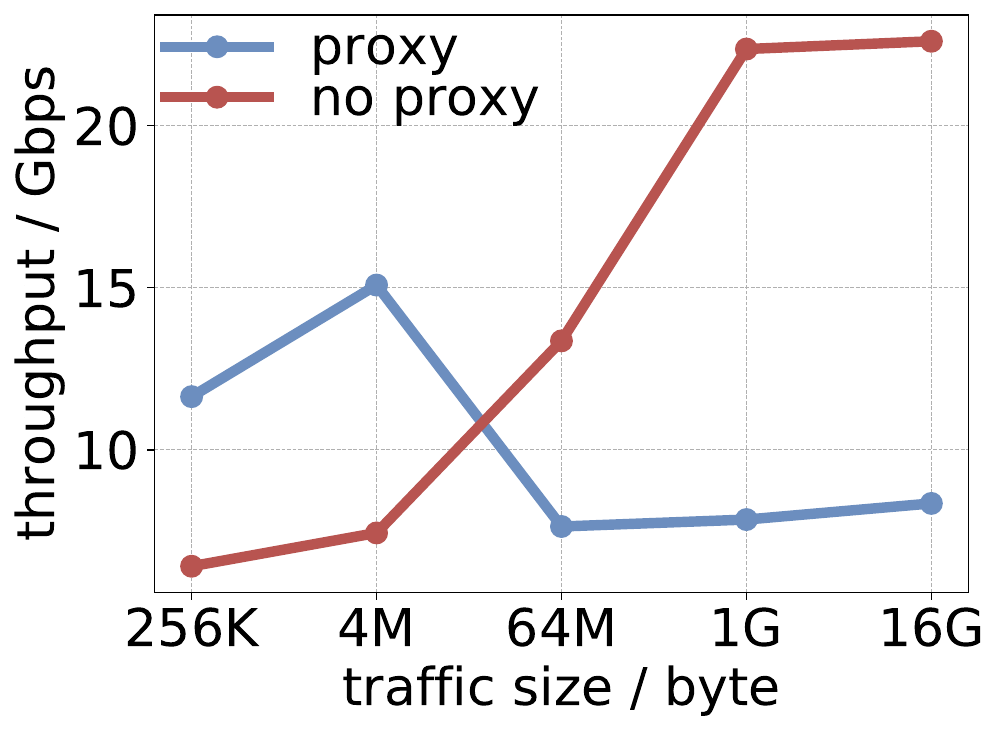}
        \label{fig:motivation:a}}
\subfloat[]{ 
        \includegraphics[width=110pt]{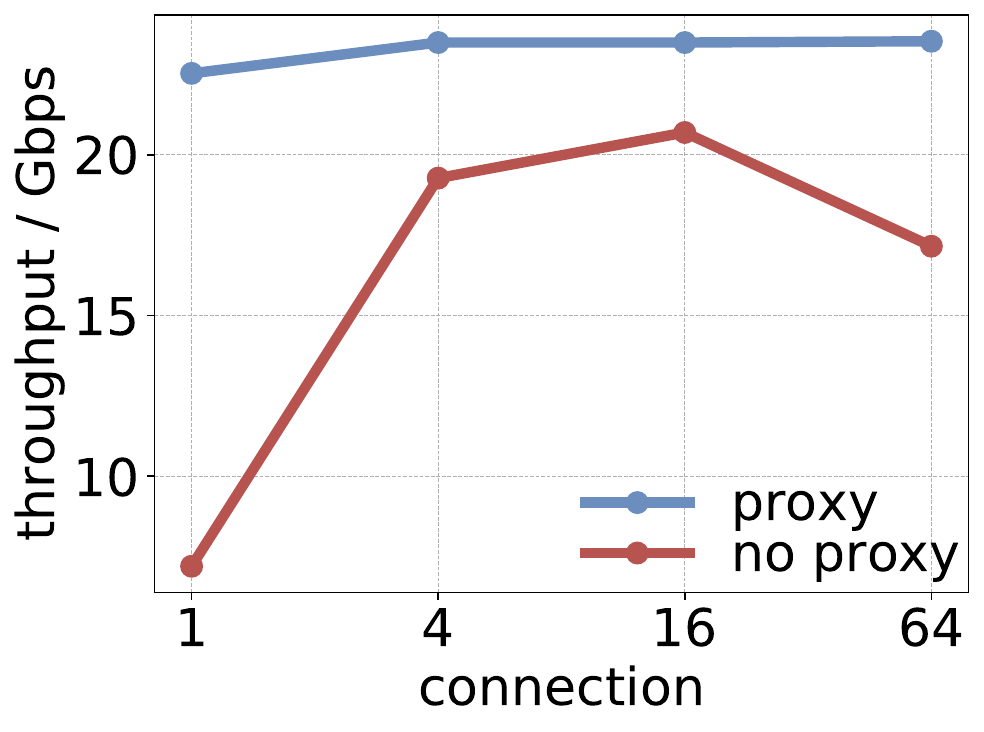}
        \label{fig:motivation:b}}
\subfloat[]{ 
        \includegraphics[width=118pt]{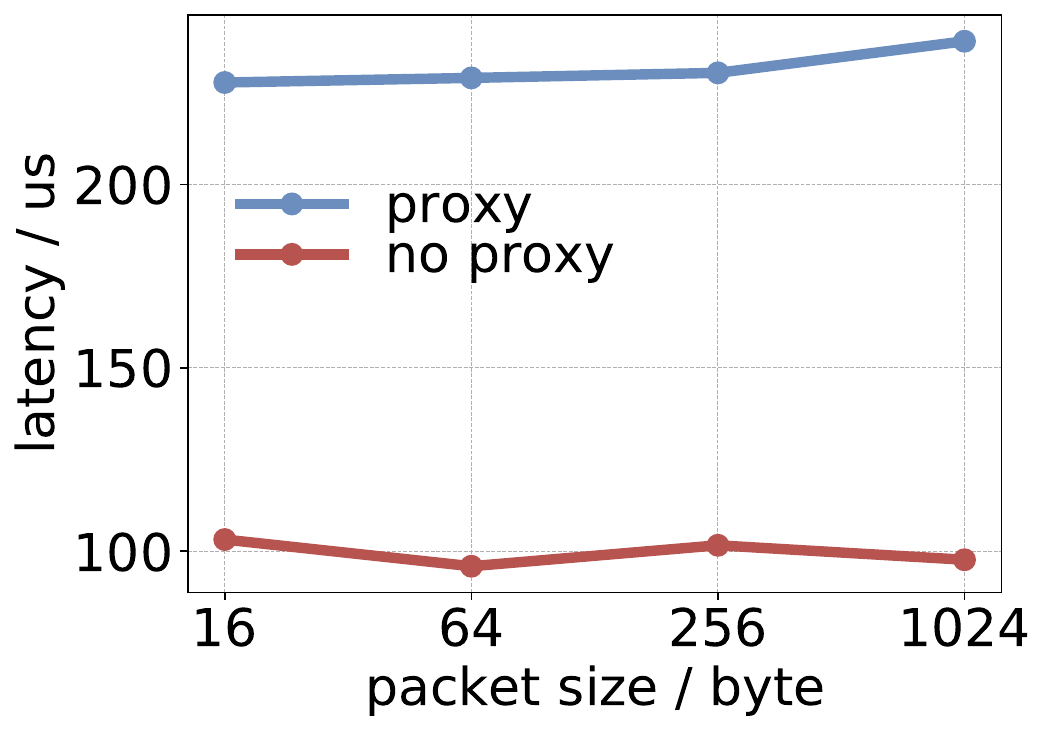}
        \label{fig:motivation:c}}
\subfloat[]{ 
        \includegraphics[width=130pt]{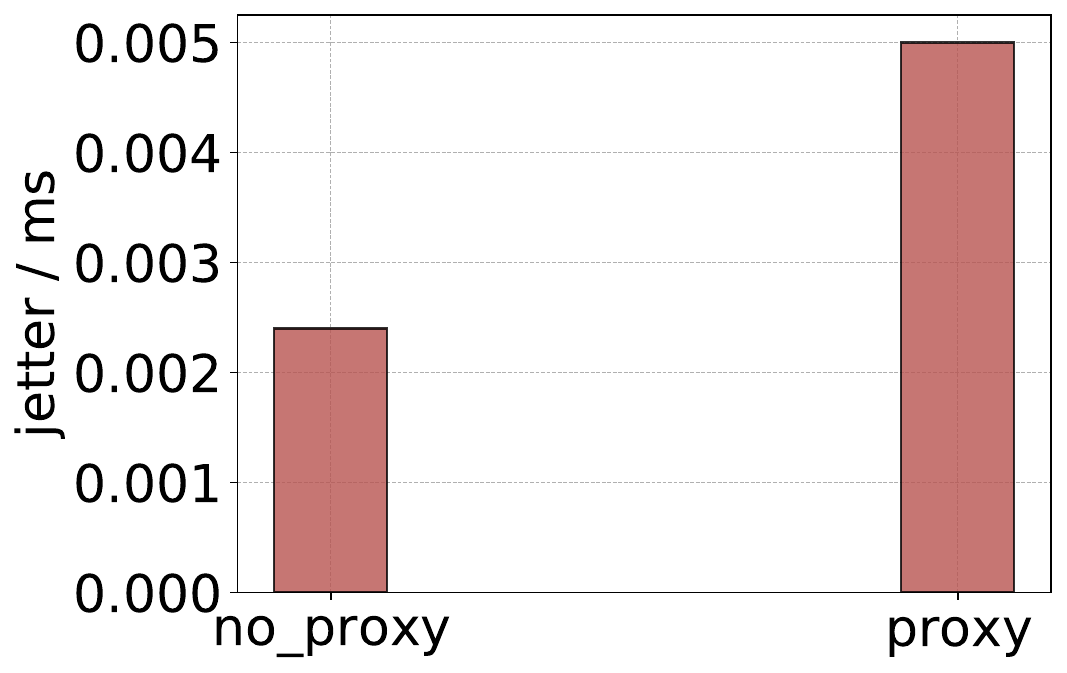}
        \label{fig:motivation:d}}                
\caption{The performance comparison inducing service mesh}
\label{fig:motivation}
\end{figure*}

\textbf{CPU overload} Sidecar proxy involves the entire network layer, which naturally includes many IO-intensive functions. Additionally, the communication of the proxy is mainly through RPC and Restful, which involve multiple computation-intensive functions such as encryption/decryption, serialization/deserialization, compression/decompression, etc. These functions overload the CPU when the sidecar proxy processes traffic. Although some research \cite{15.NSM-VPP, 17.cloud-native-gateway} has optimized the system architecture by sharing the protocol stack or implementing centralized proxies to reduce the CPU burden, CPU overload is unavoidable as the I/O bandwidth increases beyond Moore's law. In our experiment, up to 70\% of CPU resources were occupied by the sidecar proxy. Besides, CPU overload will cause performance interference for services. As shown in Figure \ref{fig:motivation:d}, the latency jitter of the proxy is far from the original service communication. In a production environment, these infrastructure functions irrelevant to business consume more than 30\% CPU resources causing high costs in the cloud data center\cite{30.infrastructure_cost}.

\begin{figure}[t]
  \centering
  \includegraphics[width=220pt ]{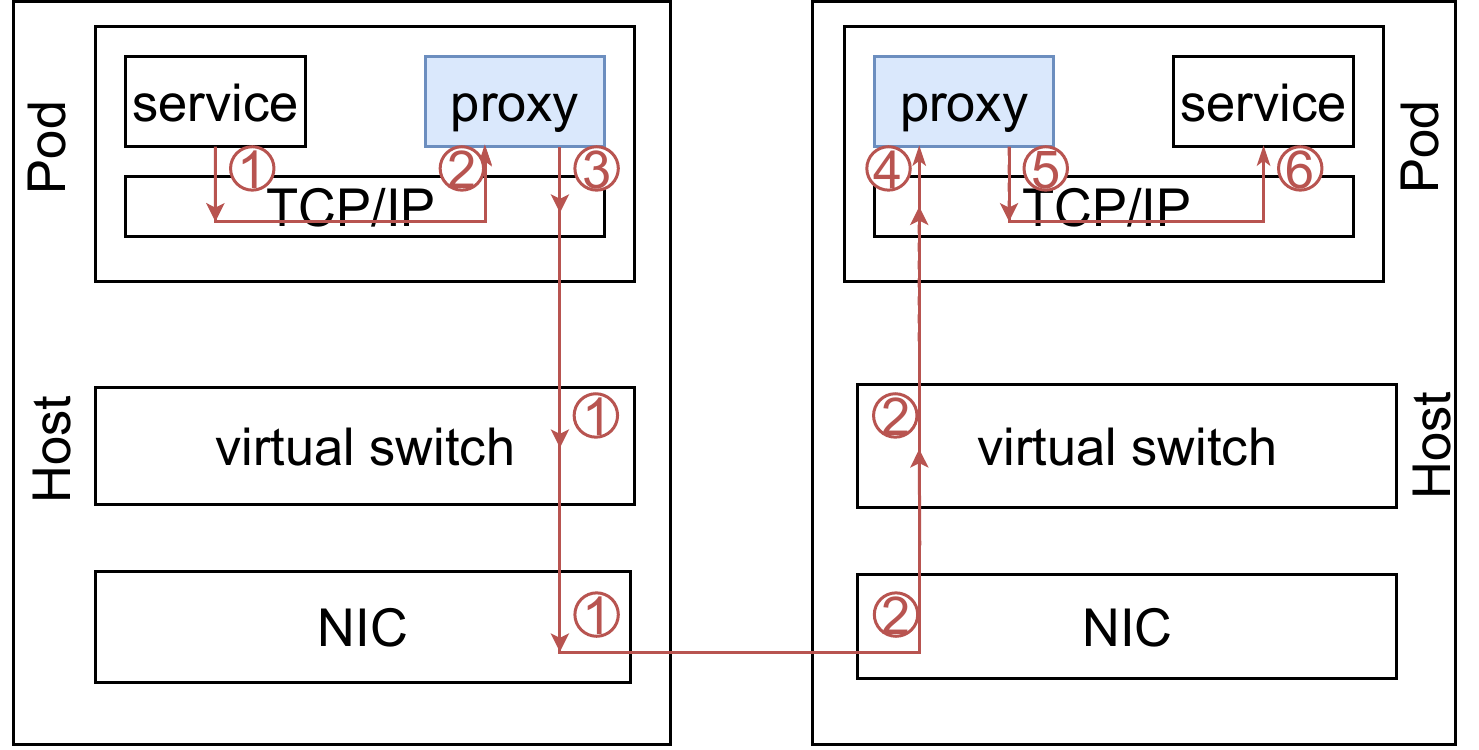}
  \caption{The communication model of service mesh.}
  \label{fig:service_mesh_communication}
\end{figure}

\textbf{Higher latency and lower throughput} According to our analysis, service communication through service mesh needs to pass through VM, host, and NIC, which creates a lengthy data path and larger software overhead. As shown in Figure \ref{fig:service_mesh_communication}, service communication includes six network stack processing, two vSwitch forwarding, and more than ten data copying operations. Additionally, the execution of the proxy also includes many system overheads such as context switches, resource competition, function calls, etc. These overheads seriously degrade communication performance and stability. As shown in Figure \ref{fig:motivation}, our results show a 70\% decrease in throughput and an increase in latency by more than 2x.

\begin{figure}[t]
  \centering
  \includegraphics[width=220pt ]{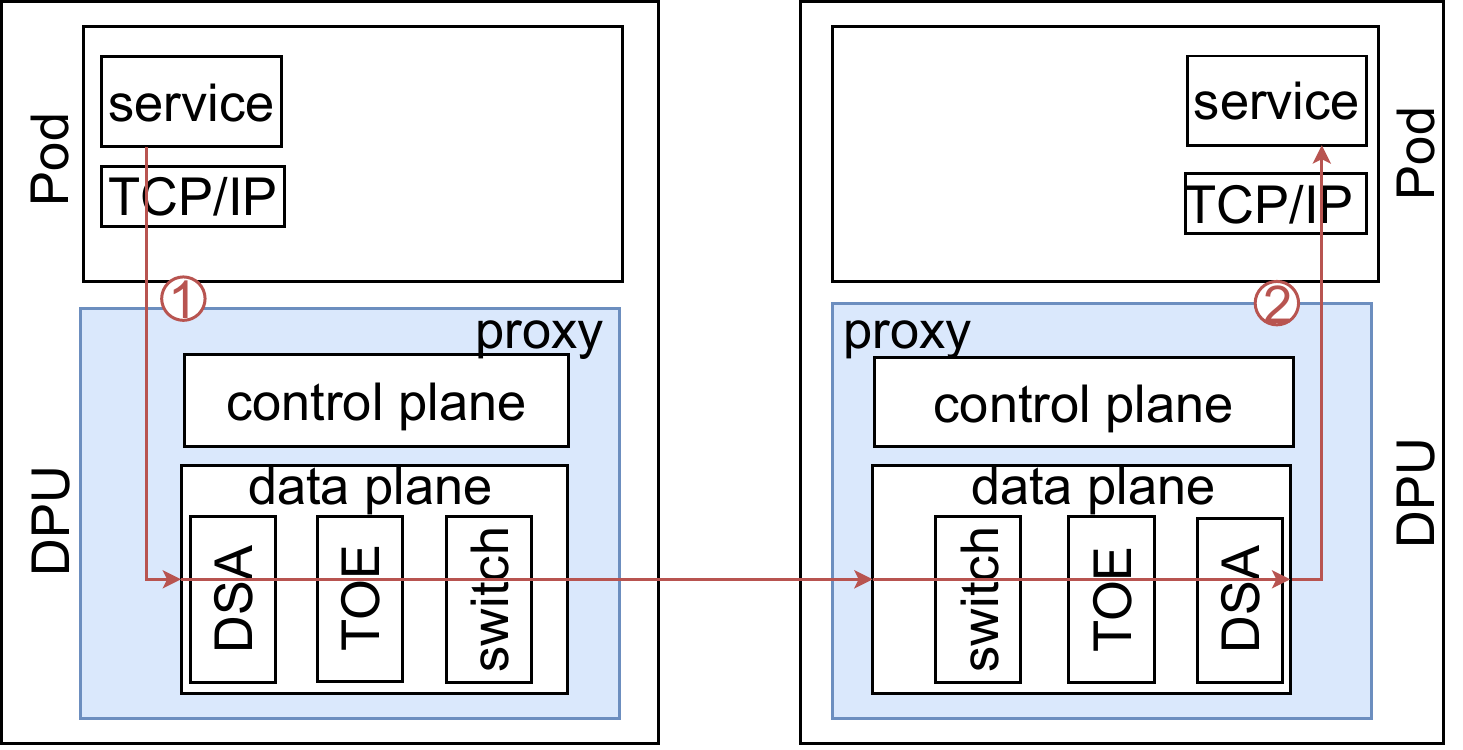}
  \caption{The communication model of FlatProxy}
  \label{fig:flatproxy_communication}
\end{figure} 

\subsection{Our Goal}
Therefore, we propose the data-centric service mesh FlatProxy, which migrates the service mesh system to the entrance of the network. As shown in Figure \ref{fig:flatproxy_communication}, deploying the service mesh at the entrance of the network can govern all communication in and out of nodes, meeting the requirements of data-centric applications. Moreover, the physical isolation between the cloud infrastructure and application reduces the consumption of CPU resources. We also aim to leverage the accelerator in the hardware platform to accelerate data processing and optimize the communication path. Ideally, our design only requires data to be copied twice between the service and the network.

\section{Hardware Service Mesh and Challenge}
In this section, we introduce the hardware platform to offload the service mesh and the challenges of implementing the service mesh in hardware. 

\subsection{Hardware Technology}
To implement FlatProxy, we have examined various hardware platforms from four aspects: interconnection, control, acceleration, and programmability.

\paragraph{Network-attached Architecture} 
The network-attached architecture implements direct attachment between NIC, storage, or accelerator \cite{63.NVMe-NIC, 64.GPUDirect}, which improves system performance by allowing communication between the network and device to bypass the CPU. Some research \cite{62.lambda-NIC,58.Lynx} has offloaded data processing to NIC, which is similar to the process. However, these architectures generally depend on specific scenarios and are not suitable for the diverse traffic transmission in the cloud.

\paragraph{SmartNIC} 
SmartNIC is a powerful hardware platform that generally has programmable logic to accelerate the computationally intensive task. Many studies have discussed the use and design of SmartNIC in different scenarios, and the primary goal of SmartNIC is to accelerate network data processing \cite{45.accelNet, 66.smartDS, 46.AccelTCP}. However, SmartNIC lacks the ability to connect different devices and does not always have the control capability to support complex control functions like limiting, circuit breaking, etc.

\paragraph{DPU} 
DPU is the evolution of SmartNIC \cite{19.fungible,18.intel,20.nvidia,21.pensando,29.yusur}. It not only accelerates data processing but also supports control capability through the embedded CPU. Meanwhile, it has a mechanism for interconnecting the accelerator, host CPU, network, and other components. The programmable switch provides smart routing to transfer traffic according to traffic identification like IP, port, URL, etc.

As shown in Table \ref{tab:summary}, DPU exhibits excellent characteristics in all four aspects. Therefore, we have chosen DPU as the hardware acceleration platform to offload the service mesh.

\begin{table}
\begin{center}
\caption{The characteristic of hardware platform (I: interconnection, C: conrtol, A: acceleration, P: programmability)}
\label{tab:summary}
% \centering
% \begin{tabular}{p{3cm}<{} p{0.6cm}<{}p{0.6cm}<{} p{0.6cm}<{} p{0.6cm}<{}}
% \begin{tabular*}{\hzise}{l|cccc}
\begin{tabular}{| l | c | c | c | c |}
\hline
% \toprule
  Hardware Platform  & I &  C  & A & P \\
%\midrule
\hline
  Network-attached    & \Checkmark     &  \XSolidBrush & \XSolidBrush & \XSolidBrush \\
\hline
  SmartNIC            & \XSolidBrush   &  \Checkmark   & \Checkmark   & \Checkmark   \\
\hline
  DPU                 & \Checkmark     &  \Checkmark   & \Checkmark   & \Checkmark   \\
\hline
  
%\bottomrule
\end{tabular}
\end{center}
\end{table}

\subsection{Data Processing Unit(DPU)}

In this section, we provide a summary of a typical DPU architecture that defines its primary features. We also present a hardware instance that represents the current form of DPU.

\subsubsection{What is DPU}
\begin{figure}
\centering     %%% not \center
\subfloat[common architecture]{ 
    \includegraphics[width=90pt]{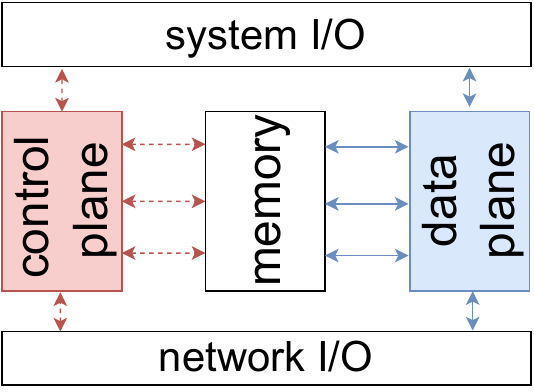}
    \label{fig:dpu-arch:a}}
\subfloat[hardware instance]{
    \includegraphics[width=120pt]{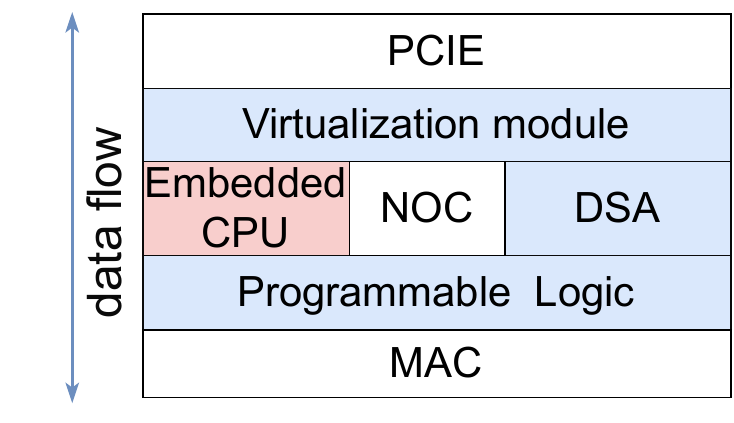}
    \label{fig:dpu-arch:b}}
\caption{DPU model (a)common architecture(b) Function model. (1) PCIe is the interface between the host and DPU.(2)The virtualization management module schedules the hardware to the virtual machine. (3) DSA includes diversified accelerators that process data of different domains; NOC is the communication bridge between different modules; the embedded CPU controls and manages the hardware. (4) The programmable logic processes data packets of different network protocols. (5)The MAC is the interface between DPU and the network.}
\label{fig:dpu-arch}
\end{figure}

As shown in Figure \ref{fig:dpu-arch:a}, DPU comprises five parts: system I/O, network I/O, control plane, data plane, and memory \cite{19.fungible,18.intel,20.nvidia,21.pensando,29.yusur}.
The system I/O serves as the communication interface between DPU and other processing units, such as CPU, FPGA, and GPU.
The network I/O includes the high-speed network interface, which enables communication between DPU and the network.
The control plane (1) manages all resources; (2) controls the data path running, and (3) allocates computing tasks to hardware.
The data plane accelerates data processing and generally includes domain-specific accelerators and high-speed programmable logic.
The memory is used to store intermediate results, device status, and configuration information.

\subsubsection{How to solve the problem using DPU}
A typical hardware implementation, as shown in Figure \ref{fig:dpu-arch:b}, includes high-performance PCIe and network interfaces, virtualization modules, programmable logic, a set of accelerators, embedded CPUs, NOCs, and more. These hardware modules support the ability to implement the service mesh.

\textbf{Abundant I/O interfaces support interconnection between hosts, networks, and other devices.} Firstly, DPU implements high-speed PCIe to communicate with the host. Generally, PCIe supports I/O virtualization by SRIOV. The VM or container can directly access the device, bypassing the host kernel by SRIOV, making virtual I/O device performance close to the physical device \cite{22.SRIOV}. Secondly, DPU implements high-speed network interfaces to communicate with the outside network. Thirdly, DPU implements PCIe switch to support interconnection with other devices like GPU. Besides, DPU implements NVMe-of to connect high-performance storage devices.

\textbf{Diverse accelerators accelerate data processing.} DPU implements various accelerators to speed up network processing. For example, TOE is used to accelerate TCP processing; Regex engine is used to accelerate character string matching; and other accelerators can also accelerate other computing-intensive operations like encryption/decryption, serialization/deserialization, compression/decompression, etc.

\textbf{Programmable logic supports flexible data processing and smart routing.} DPU implements programmable logic like vSwitch and NP. VSwitch supports basic routing below L4 and advanced functions such as VLAN, VXLAN, NAT, and TC. NP (network processor) can implement more complex functions such as HTTP parsing, L7 routing, and load balancing. These functions can be executed according to identifiers such as IP, port, URL, etc.

\textbf{General-purpose processors support the implementation of control functions.} DPU integrates general-purpose processors such as embedded CPUs, which are essential for implementing control functions. For example, they can manage TCP connections and control the flow of data processing. Moreover, general-purpose processors enable development to implement the control plane on DPU, reducing performance interference and security issues with complete isolation between applications and cloud infrastructure.

In addition, DPU builds a complete framework for application security and monitoring\cite{72.zero}. For example, it supports hardware for "zero-trust" security architecture. It can observe data flow through statistics, logs, and trace connections.

\subsection{Implementation based on Embedded CPU}
Some prior work has implemented infrastructure functions using the embedded CPU of the device \cite{62.lambda-NIC, 68.DPFS}. In this section, we intend to learn from this method to implement service governance on the embedded CPU provided by BlueField2 (BF2).

\begin{figure}[t]
  \centering
  \includegraphics[width=200pt ]{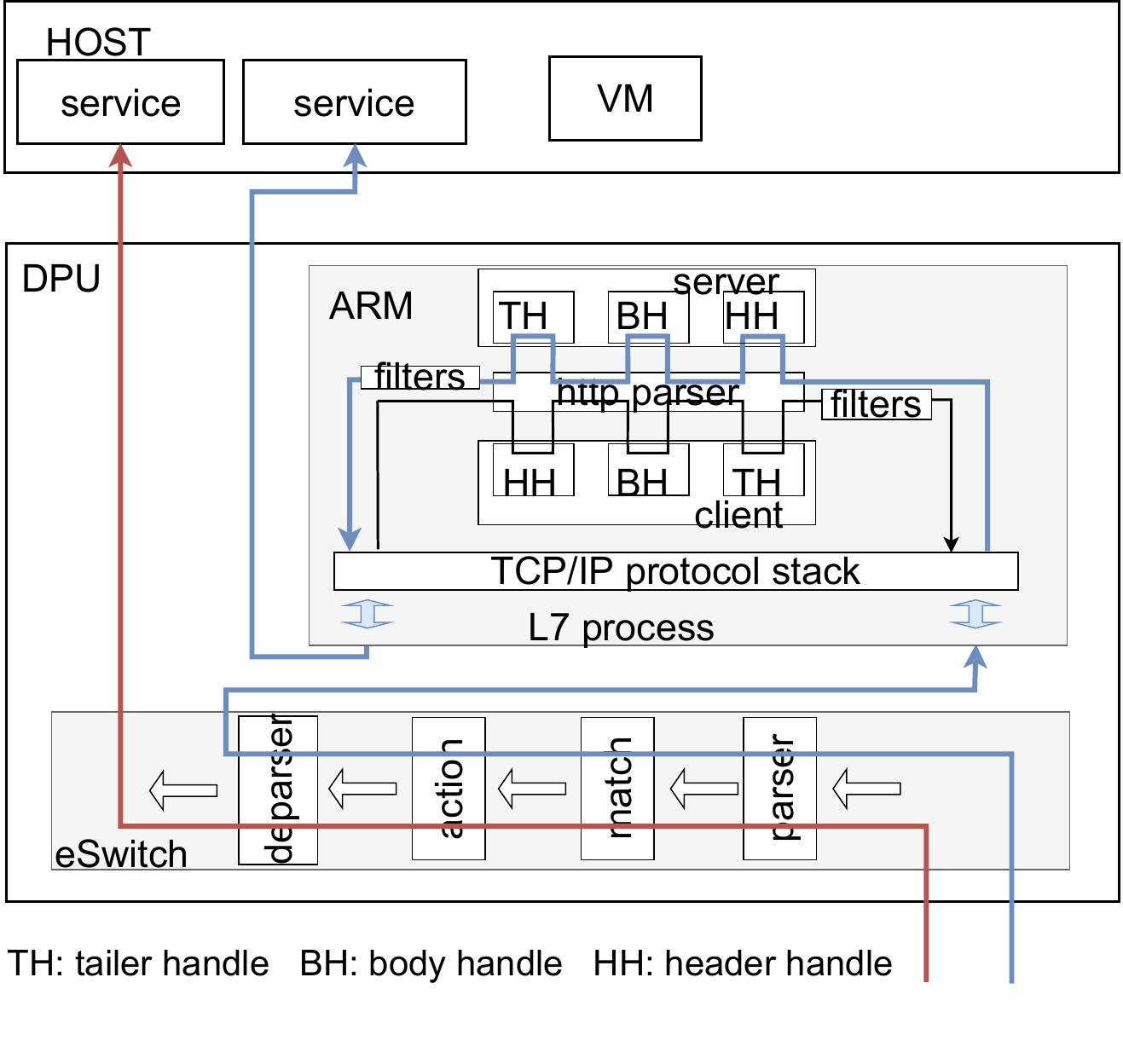}
  \caption{The architecture of sidecar prosy with BF2}
  \label{fig:bg2-arch}
\end{figure}

As shown in Figure \ref{fig:bg2-arch}, some service governance functions (L3/L4), like routing, can be implemented on the eSwitch, and some advanced functions of the application layer (L7) can only be implemented on the ARM. When traffic arrives at the DPU, it will be forwarded to the ARM through the eSwitch if the traffic needs to be handled by these L7 functions. The processed traffic is then sent to the service through the eSwitch. However, if the traffic only needs to be processed by some function that is implemented on the eSwitch, it will be directly sent to the service. This way, we can implement service governance on the DPU.

\begin{figure}[t]
\centering
\subfloat[The latency using eSwitch]{
    \includegraphics[width=0.45\linewidth]{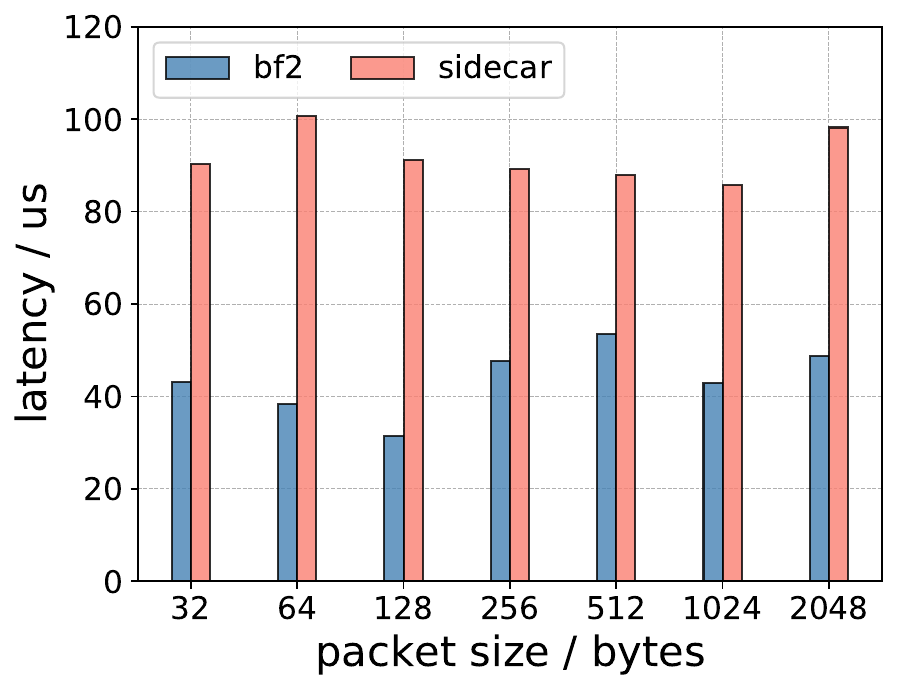}  
    \label{fig:bf2-latency}
}
\subfloat[The throughput using eSwitch]{
    \includegraphics[width=0.45\linewidth]{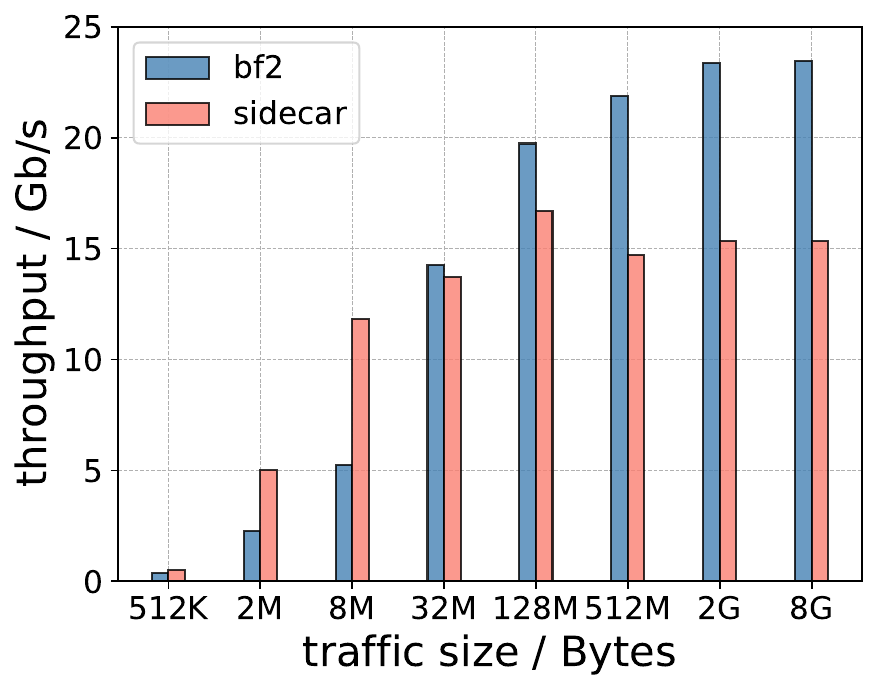}  
     \label{fig:bf2-throughput}
}

\subfloat[The respond rate of sidecar proxy]{
     \includegraphics[width=0.5\linewidth]{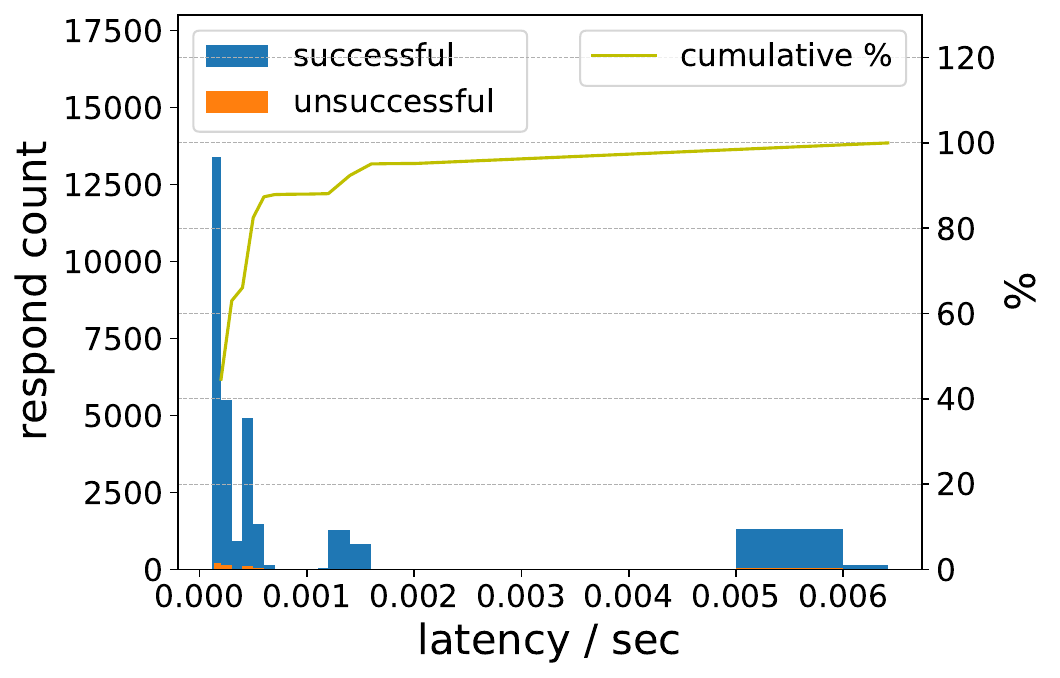}  
     \label{fig:sidecar-l7}
}
\subfloat[the respond rate of bf2 using ARM]{
     \includegraphics[width=0.5\linewidth]{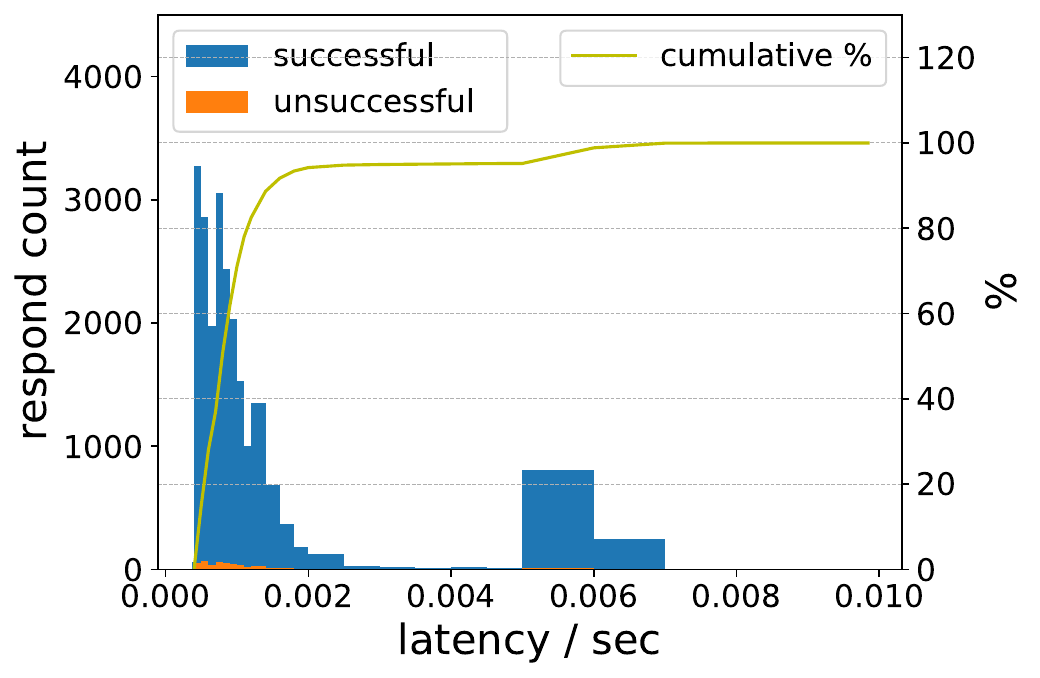}  
     \label{fig:bf2-l7}
}

\caption{The performance comparison between sidecar and bf2}
\label{fig:bf2_impact}
\end{figure}

However, it is challenging to meet the performance requirements. The results shown in Figure \ref{fig:bf2_impact} indicate that BF2 improves the performance of the functions implemented by the eSwitch, but it degrades the performance of the functions implemented on the ARM. In fact, the eSwitch can only implement a few functions about the service mesh, which is not enough to support the entire system of the service mesh. To implement the whole system of service governance, we analyzed the performance of Envoy \cite{6.envoy}, a popular implementation of a sidecar proxy running on the ARM to implement service governance. We observed that the reasons for performance degradation include three points: (1) the embedded ARM core does not have enough data processing capability; (2) the execution process of Envoy is serial, which causes data processing to be blocked by other functions such as statistics; (3) the implementation on the ARM is similar to Envoy on the host, which creates software overhead, such as context switching, data copying, CPU competition, etc.

Therefore, we propose FlatProxy to overcome these problems and achieve a balance between performance and flexibility. FlatProxy adapts the software/hardware co-design to rebuild the service mesh based on the common architecture of the DPU.

\subsection{Chanllenge}
Implementing FlatProxy involves three challenges involving traffic processing, function mapping, and communication isolation. We introduce these challenges as follows.

\textbf{Higher Traffic Flow} Compared to a sidecar proxy, FlatProxy not only needs to process traffic in and out of the node but also needs to process traffic between services in the node. Theoretically, the peak value of traffic is more than the network bandwidth. In addition, multiple computing-intensive functions are also a barrier to implementing higher processing bandwidth. To solve this problem, we design a hierarchical processing architecture that applies a pipeline structure to process the traffic in L2, L3, and L4 and applies a multi-core structure to process the traffic in L7. Besides, some computing-intensive processing, like encryption/decryption for payload, will be executed by DSA. By hierarchical processing, FlatProxy processes traffic by traffic category, which benefits from leveraging the unique feature of various hardware to improve the system performance.

\textbf{Handling Diverse Function} Service mesh involves diverse functions, including four aspects: service discovery, traffic control, observability, and security. These functions have different features, like control-intensive, computing-intensive, I/O-intensive, etc. Meanwhile, functions can be dynamically made up to function chains according to the type of data. These characteristics mean FlatProxy needs abundant processing functions and a flexible way to organize these functions. These requirements are easy for software, but how to implement an extensible and flexible hardware-accelerated data processing platform is not a simple thing. To address this challenge, we firstly classify functions into configuration, control, and data processing. And we only accelerate these functions about data processing because it generally is the performance bottleneck. Secondly, we propose a hardware thread model that extends the match-action mechanism \cite{34.p4} to implement the flexible software-defined data processing flow.

\textbf{Multi-tenancy Performance and Security Isolation} The multi-tenant scenario has the demand for performance and security isolation. Many researchers solve the challenge in the host CPU, but it is unavoidable that tenancies affect each other due to software operation, like schedule, context switch, interrupt, resource contention, etc. In our design, the natural isolation between service and infrastructure reduces the software overhead and the security vulnerabilities. To further solve this problem, we leverage SRIOV and socket direct to optimize the communication path between service and FlatProxy, which reduces the participation of the host and protocol stack in the container to improve the performance. Meanwhile, using SR-IOV to implement the memory isolation ensures communication security. By this method, FlatProxy improves the performance and communication stability.

\section{Design and Implementation}
\subsection{Design overview} \label{design-model}

\begin{figure}[t]
  \centering
  \includegraphics[width=200pt]{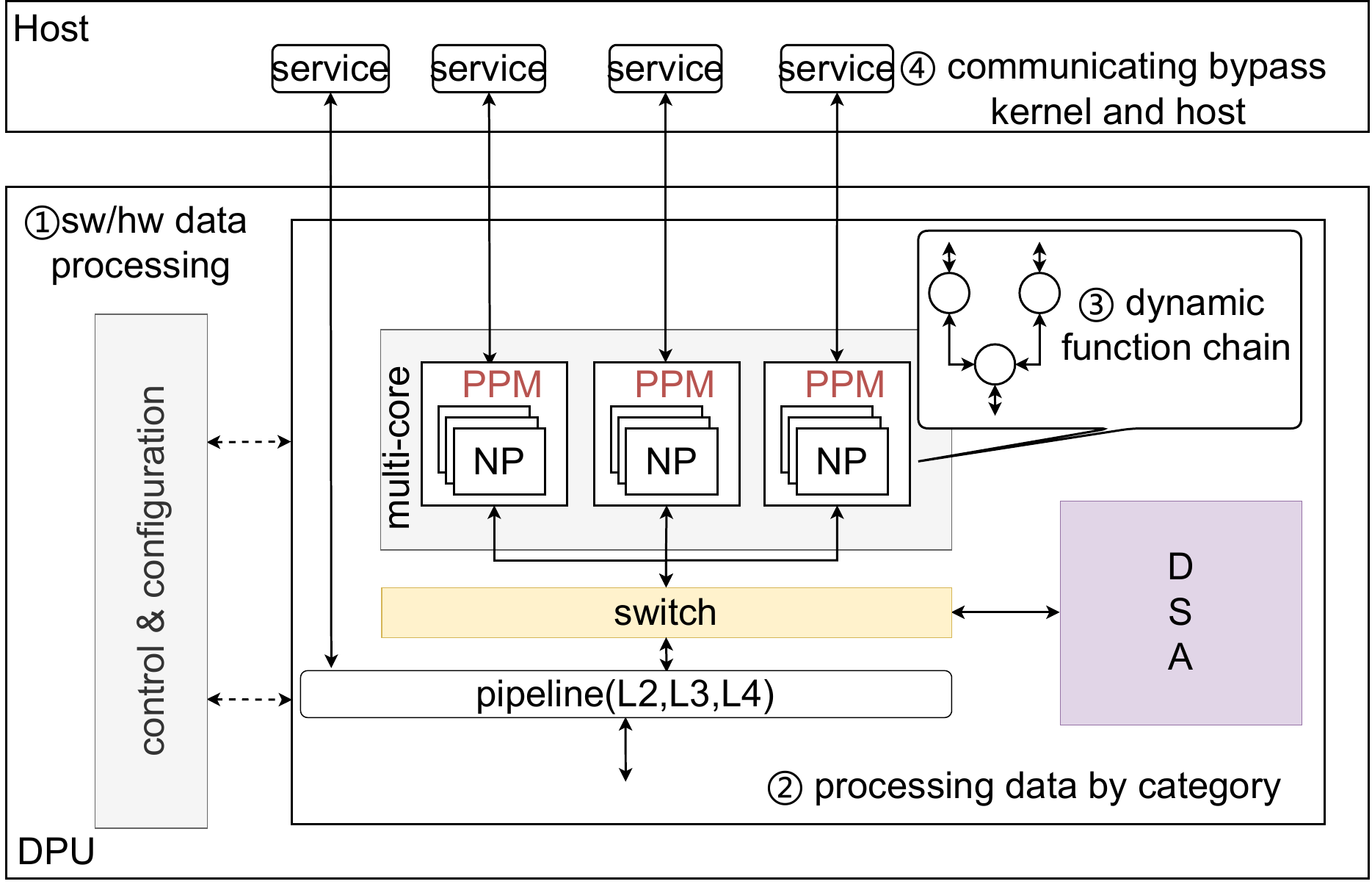}
  \caption{The Architecture of FlatProxy}
  \label{fig:flatproxy_arch}
\end{figure}

To implement service governance for large-scale data-centric applications in multi-tenancy clouds, we designed FlatProxy based on DPU. Figure \ref{fig:flatproxy_arch} shows the high-level architecture. We leveraged a multi-pronged approach to balance performance and flexibility.

\textbf{Hardware and Software Co-Design}
We classified service governance features into data processing, control, and configuration. FlatProxy implements data processing using special hardware and a proxy as the control plane on the embedded CPU. The proxy achieves control features like connection management, control and configuration of data processing, and communication with the cluster's control center.

\textbf{Hierarchical Processing Architecture}
FlatProxy improves data processing performance through a hierarchical processing architecture that processes data by traffic category. Functions in L2, L3, and L4 can be implemented by a pipeline to provide maximum bandwidth. The function in L7 can be processed using a multi-core architecture to balance performance and flexibility. Additionally, computing-intensive tasks are processed using accelerators.

\textbf{Hardware Multi-Thread Model}
FlatProxy proposes a multi-thread model to achieve dynamic function combinations. The model includes multiple layers, and each layer includes a set of packet processing modules (PPMs) that only communicate with adjacent layer PPMs. We extended the match-action mechanism to achieve diverse data processing functions and dynamic combinations.

\textbf{Communication Path Optimization}
FlatProxy optimizes communication paths by bypassing the host and container protocol stack using SRIOV and socket direct to enable direct communication. Communication is isolated using SRIOV to ensure security in multi-tenancy scenarios.

\subsection{Hardware and Software Co-design} \label{system-architecture}
In general, the system performance bottleneck is due to I/O intensive and computation intensive functions. For example, executing I/O intensive functions like routing and filtering can degrade performance due to frequent data transmission between VM, host, and NIC. Therefore, we leverage hardware and software co-design to reduce data transmission and accelerate data processing. In the top design, FlatProxy includes three parts.

\textbf{Configuration Plane:} This plane needs to receive the configuration file from the cloud control center and resolve the file into control rules and configuration information. This information includes the execution order of functions, routing rules, and checking rules that determine how to process traffic. An automatic configuration method can reference the current service mesh implementation, and we implement it based on Envoy, an open-source proxy.

\begin{figure}[t]
\centering
\includegraphics[width=220pt]{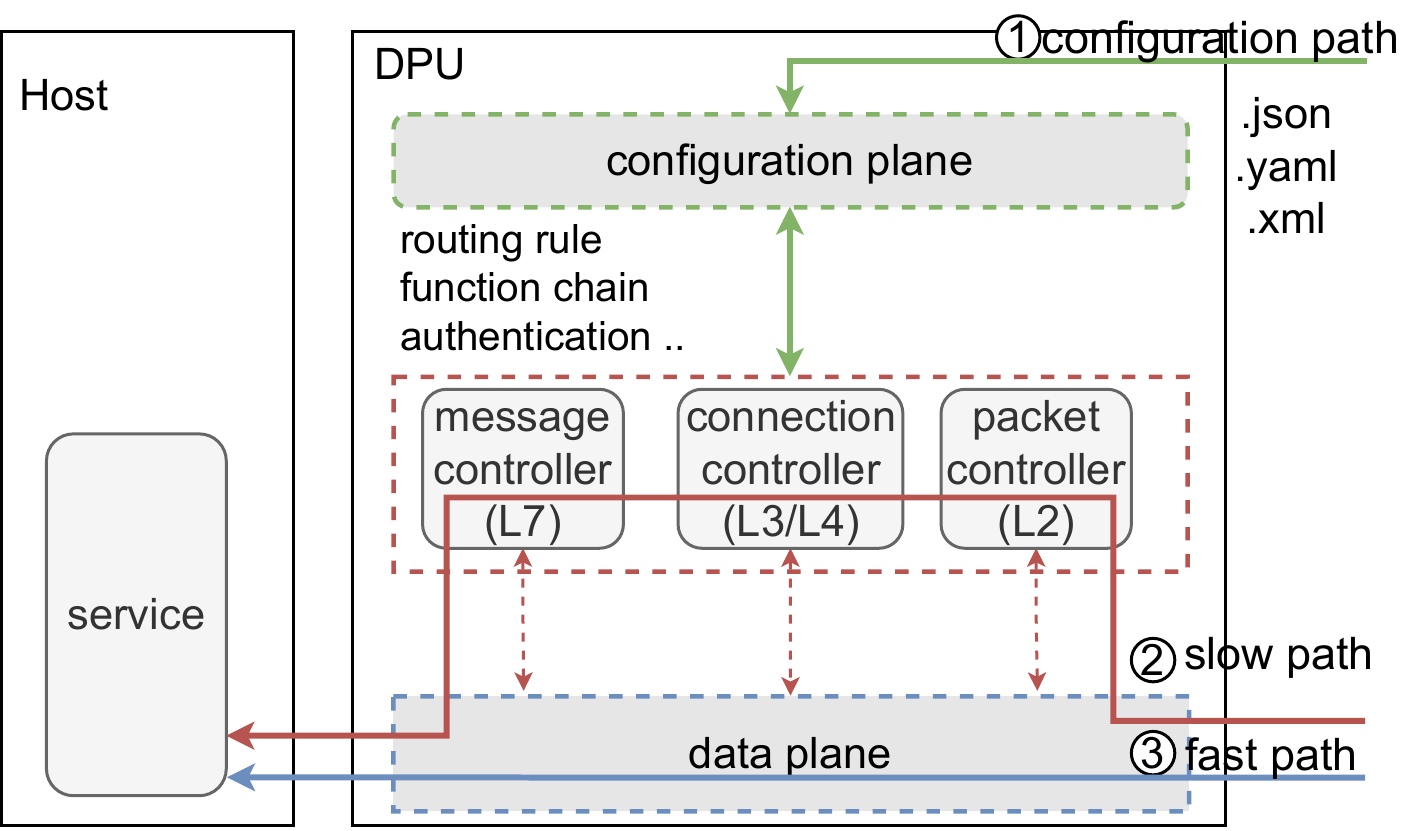}
\caption{The architecture of software and hardware co-design}
\label{fig:flatproxy_sw_hw}
\end{figure}

\textbf{Control Plane (Slow Path):} The control plane needs to implement multiple functions. Firstly, it needs to implement control features like connection management. Secondly, it needs to implement the slow path, which processes new connections and distributes rules to the data plane. Thirdly, it needs to communicate with the cloud control center to obtain control rules and send endpoint statuses, such as connection status, service status, and hardware status, to the cloud control center. To enhance scalability, we propose a layer-wise control model for efficient deployment based on network hierarchy. The model includes multiple controllers, such as the OVS controller in L2, the connection controller in L3/L4, and the message controller in L7.

\textbf{Data Plane (Fast Path):} The data plane is a fast path to process traffic. It improves performance by reducing data transmission, and the hardware can accelerate data processing. Generally, it sends new connections and other traffic it cannot process to the control plane, and when it receives the configuration and control rules, it constructs the data processing chains. Then, data from the network or service is processed through a set of processing units.

The typical way to apply FlatProxy is shown in Figure \ref{fig:flatproxy_sw_hw}. First, the configuration plane receives configuration information from the control center (such as Istio), resolves the configuration file, and sends it to the control plane. Second, the slow path processes the first packet, new connection, and other control functions. Third, the fast path is constructed by a set of special hardware. Generally, the fast path processes traffic according to identifiers like IP, port, URL, etc.

\subsection {Hierarchical Processing Architecture}

The protocol in the lower layers (L2, L3, L4) is generally unchanging and easy to be resolved by hardware. Meanwhile, service mesh functions like forwarding, filtering, and modifying the header are simple and require less state. Therefore, we adapt the Reconfigurable Match Tables (RMT) pipeline architecture \cite{73.rmt} to accelerate data processing in the lower layer. However, protocol processing and service governance in application layers (L7) are variable and complex. For example, routing may depend on the URL, which is a variable length string, while operations in this layer may need to process the payload, such as serialization/deserialization. The simple pipeline is insufficient for these requirements, so we adapt a multi-core architecture to strike a balance between performance and flexibility. In our prototype, we adapt the Network Processor (NP) as the basic processing unit, which supports special instruction programming. Additionally, computing-intensive operations such as TSO and TRO in L3/L4 and encryption/decryption of payload require special hardware to accelerate processing.

\begin{figure}[t]
\centering
\includegraphics[width=220pt ]{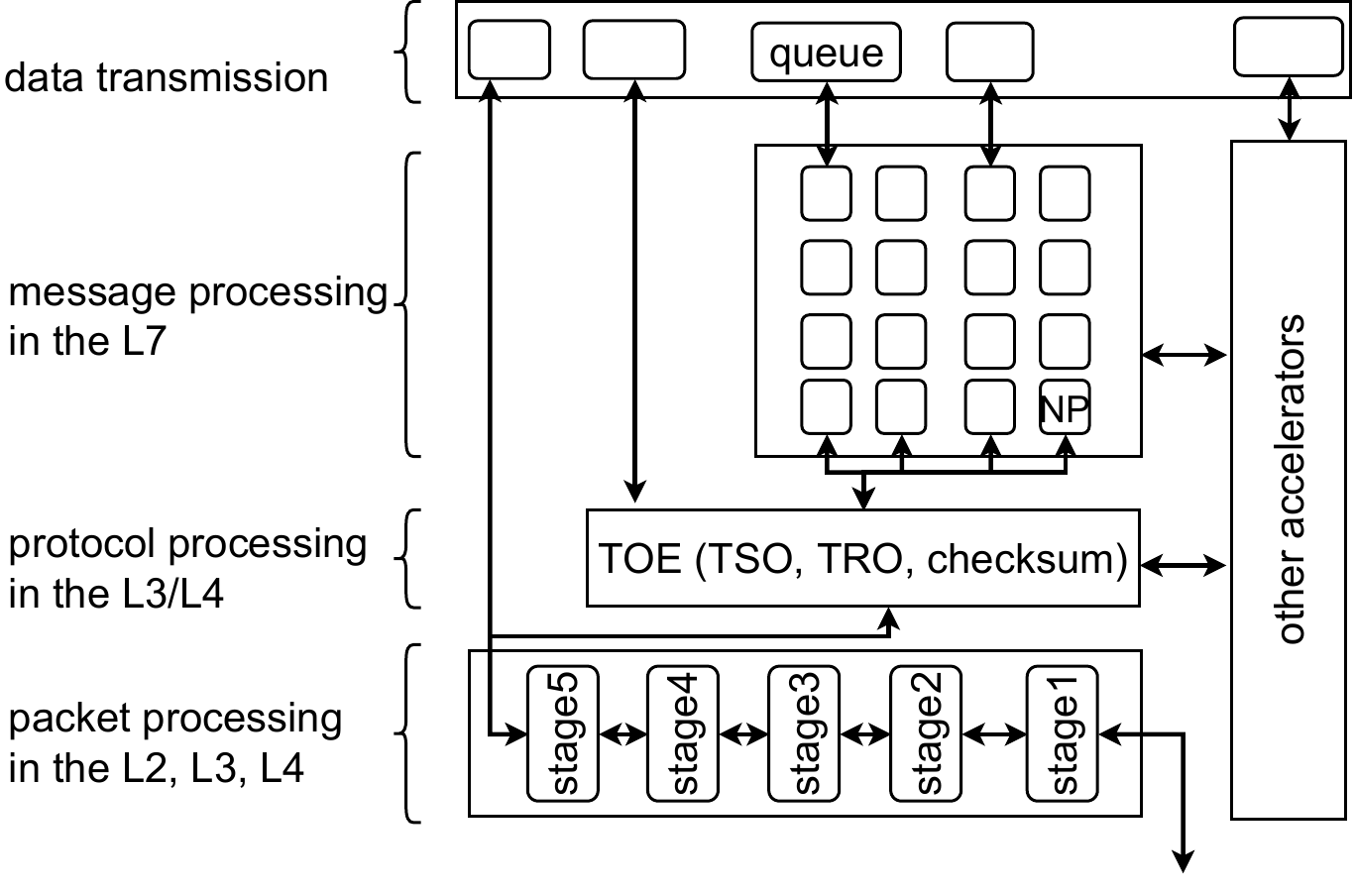}
\caption{Hierarchical processing architecture}
\label{fig:flatproxy_hierarchy}
\end{figure}

In summary, the hierarchical architecture that processes traffic by its protocol characteristics has three major benefits.

\textbf{Higher Performance}: In real systems, traffic types are diverse. Our design can classify traffic according to identifiers and process it using appropriate hardware, which improves parallelism and processing latency. Besides, suitable data also fully utilizes hardware capability.

\textbf{Simplified Architecture Design}: Classification separates service governance into several subproblems, reducing system complexity. Existing technologies such as OVS, TOE, NP, etc., can be adapted to implement the full system. The design supports flexible interconnection, and Panic proposes a high-performance interconnection method \cite{51.panic}. In the future, an efficient NoC will be designed to improve performance.

\textbf{Great Scalability}: Service mesh is continually evolving, and new features are used in different scenarios. Good scalability benefits its use in broader domains. Our design provides flexible programmable capabilities that serve as a natural interface for adding new features. Additionally, the loosely-coupled interconnection design makes it easy to add new hardware to this architecture.

\subsection{Hardware Multi-thread Model}
\begin{figure}[t]
  \centering
  \includegraphics[width=120pt ]{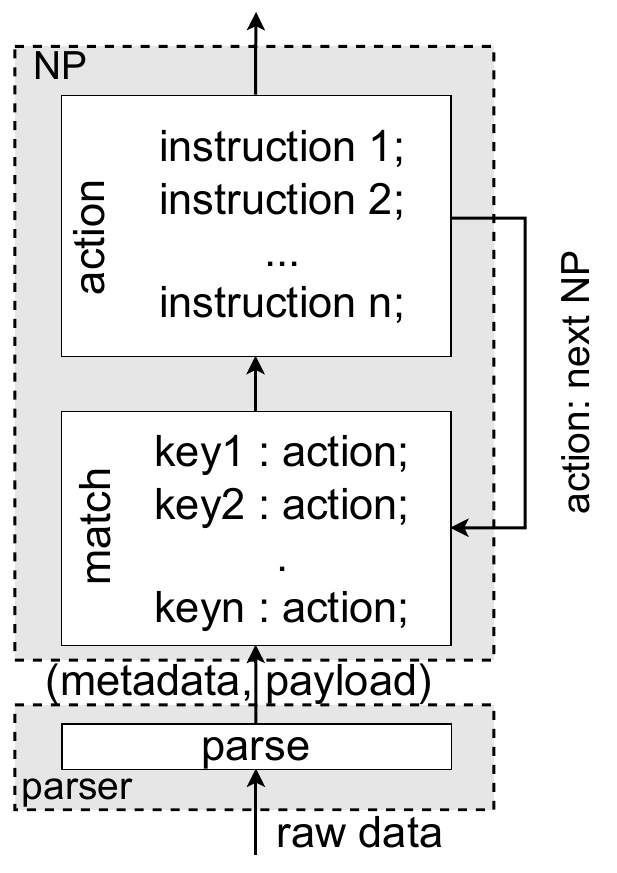}
  \caption{The hardware multi-thread model }
  \label{fig:multi_thread}
\end{figure}

The function of service mesh involves the entire network stack, and each layer's function only communicates with the adjacent layers. Therefore, the packet processing module (PPM) is only interconnected with these adjacent PPMs. The PPM is the basic logic unit of our model to process traffic. It extends the match-action mechanism as follows to form the dynamic function combination shown in figure \ref{fig:multi_thread}.

$$ <parser, (match, action)^*> $$

\begin{itemize}
    \item \textbf{Parser}: In this stage, raw data is parsed into payload and metadata. The metadata, composed of protocol header fields, includes a set of keywords used to choose an action.
    \item \textbf{Match}: In this stage, the metadata is used as keywords to select the action from the action list, and then the action instructions are loaded to the action module.
    \item \textbf{Action}: The action module processes data and sends the modified metadata and payload to the next PPM, DSA, or match module to trigger the match-action processing. The PPM can be user-customized and compatible with the existing interface like P4\cite{34.p4}.
\end{itemize}

We implement the model based on the hierarchical processing architecture. As shown in figure \ref{fig:flatproxy_arch}, the low layer PPM is implemented based on the RMT pipeline, and the application layer PPM is implemented based on the multi-core architecture. Unlike the RTC (run-to-complete) model, PPM limits each NP to execute only one function like routing, and several NPs can be dynamically constructed as function chains to process traffic. By doing so, we can achieve a flow structure to construct the function chain and improve performance. An example of HTTP routing is shown in figure \ref{fig:flatproxy_multi_thread}, and its process flow is as follows:

\begin{figure}[t]
  \centering
  \includegraphics[width=240pt ]{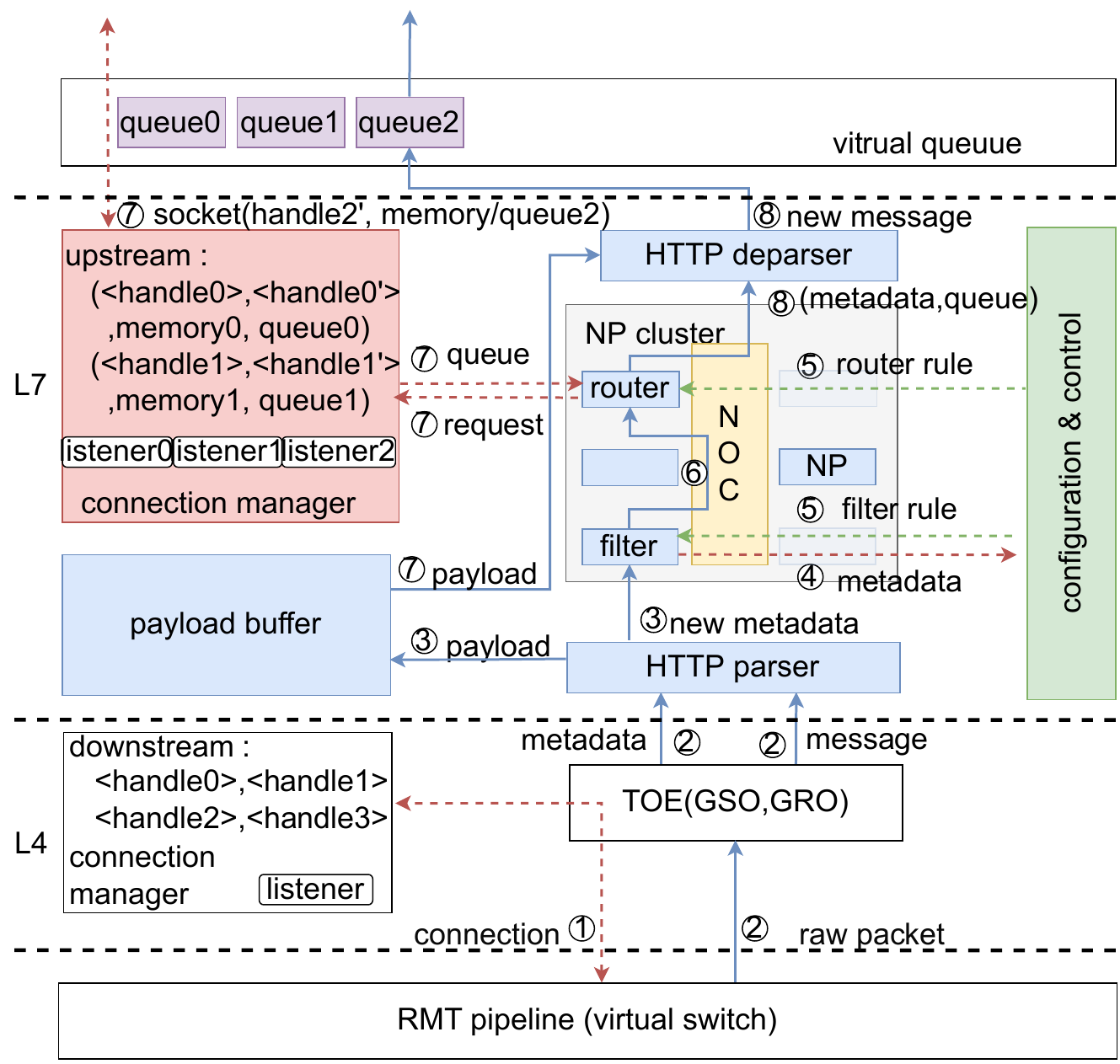}
  \caption{The example of HTTP routing}
  \label{fig:flatproxy_multi_thread}
\end{figure}

\begin{algorithm}
\caption{HTTP Routing Process}\label{alg:routing}
\begin{algorithmic}[1]
\REQUIRE $Metadata$
\ENSURE $Routing action$
\STATE $key = Key(metadata.dip, metadata.dport)$
\IF {$Listner[Hash(key)] == 1$} 
    \IF {$match(Hash(metadata.url.path)) == 1$}
        \STATE $key= Key(metadata.sip,\ metadata.sport, $
        \STATE \qquad $metadata.dip, metadata.dport)$
        \IF {$Queue[Hash(key)] == 0$}
            \STATE $endpoint = load\_balacing(metadata)$
            \STATE $Queue[Hash(key)] = connection(endpoint)$
        \ENDIF
        \STATE $metadata.queue = Queue[Hash(key)]$
        \STATE $Update\ metadata$
        \STATE $Send\ metadata\ to\ http\ deparser/DSA$
        \STATE $initial\ metadata$
    \ELSE
        \STATE $initial\ metadata$
        \STATE $break$
    \ENDIF
\ELSE
    \STATE $initial\ metadata$
    \STATE $break$
\ENDIF
\end{algorithmic}
\end{algorithm}
\begin{enumerate}
    \item TOE gets the connection information and establishes a connection with the service caller.
    \item TOE parses and recombines raw packets into a message and metadata, including IP, port, protocol type, etc. Then, the output data is sent to the processing module in the L7.
    \item HTTP parser parses the message into new metadata, including the original metadata and HTTP header fields and sends metadata to filter processing module. Meanwhile, HTTP parser caches the body payload to the buffer.
    \item The filter executes the filter rule according to metadata. If there is no rule for this request, it sends the metadata to the configuration \& control module.
    \item The configuration \& control module distributes the filter rule and router rule to the match-action logic (NP cluster).
    \item When the filter has the rule for the request, it drops or forwards to the next processing module. In this example, it sends the metadata to the router.
    \item The algorithm of the router is shown in algorithm \ref{alg:routing}, which queries the target queue to send the message. If the queue is not allocated, it requests the connection management module to establish a new connection with the service. The connection manager establishes a connection with the entity of service and binds the virtualization queue on the DPU with the socket memory in the host.
    \item The data is routed to the HTTP deparser, which encapsulates the modified header field and body data into an HTTP message and sends it to the target queue.
\end{enumerate}

\subsection {Communication Path Optimization} \label{communication_opt}

The container kernel bypass \cite{74.vpp, 75.slim} implements a multi-client access mechanism by a shared data forwarding plane, and containers communicate with the data forwarding plane by asynchronous event-notification, which may cause security problems. Additionally, when applications are deployed via serverless architecture, the short-lived execution of services or functions can result in significant performance loss due to the asynchronous event-notification.

\begin{figure}[t]
\centering
\includegraphics[width=220pt]{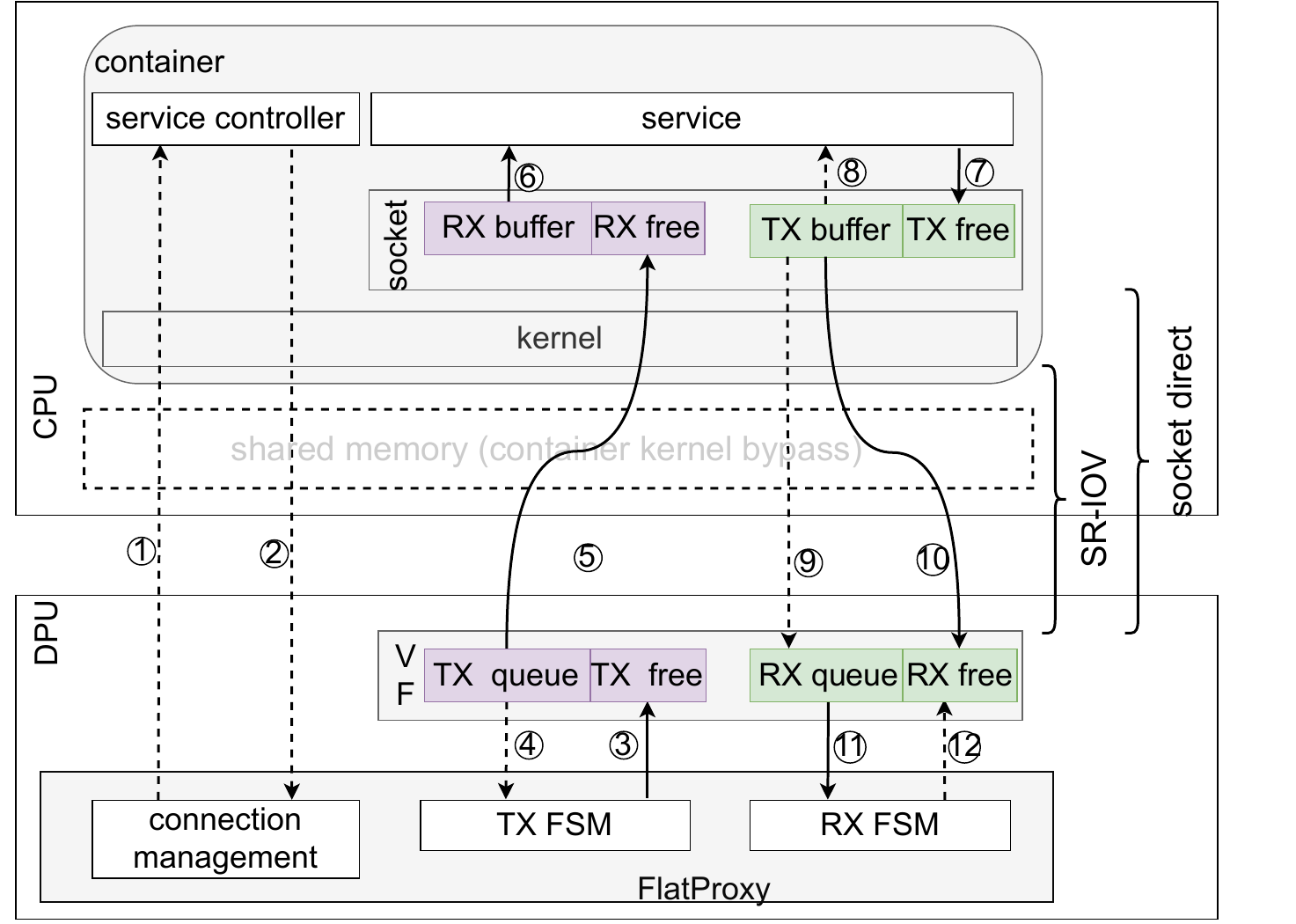}
\caption{The architecture of communication optimization}
\label{fig:communication_arch}
\end{figure}

Therefore, we leveraged the technology of SRIOV and socket direct to optimize the communication path between services and FlatProxy to improve performance and security. As shown in figure \ref{fig:communication_arch}, the container accesses hardware through SRIOV to bypass the host, while services in the container can communicate with FlatProxy through socket direct to bypass the container kernel. Compared to traditional container kernel bypassing, our optimization implements isolation of communication in the host through DMA remapping supported by SRIOV. Additionally, our optimization reduces the event wait, which is beneficial for improving communication performance. The communication process includes three parts as follows.

\textbf{Memory Binding}: When FlatProxy establishes a new connection with the service, the connection management module \textcircled{1} sends a request to the service controller via PCIe. The service controller then allocates a pair of memory blocks as the RX/TX ring and sends the allocated memory address to the connection management module \textcircled{2}. The connection management module then binds the address with the virtualization queue allocated when the connection is established.

\textbf{Transmitting Path}: Data transmission relies on the PCIe DMA initialized by the connection management module. After memory address binding with the virtualization queue, the memory address is sent to DMA when the data \textcircled{3} arrives in the virtualization queue. Subsequently, the data in the queue \textcircled{5} is sent to the socket memory by DMA without needing to notify the host using a doorbell. When the data arrives in memory, the event triggers service, which then \textcircled{6} fetches the data.

\textbf{Receiving Path}: The receiving path is more complex than the TX path. Service \textcircled{7} writes data into the TX buffer via the socket, and when the free address is \textcircled{8} returned to the service, the driver \textcircled{9} notifies the DMA to fetch the data. When the DMA gets the data address, it \textcircled{10}fetches data from the TX buffer to the Rx queue. After data arrives in the DPU, FlatProxy \textcircled{11}fetches the data and \textcircled{12} releases the RX queue.

\section{Evaluation \& Analysis}
In this section, we evaluate the FlatProxy performance in terms of throughput, latency and concurrency, which are the main evaluation indices of typical cloud data center applications. Additionally, we evaluate these performance features of FlatProxy in different network layers.

\subsection{Methodology}

We introduce the experimental setup, including the test system, workload, and benchmark systems.

\paragraph{Test System}

We built an end-to-end testing system for experimental evaluation consisting of two servers, each equipped with a DPU, and interconnected using 25G fiber optic cables. The server CPUs are Intel® Xeon® Gold 5218R CPUs @ 2.10GHz, with a total of 80 cores. The DPUs are self-developed K2-Pro DPUs by YUSUR, whose prototype verification platform is based on the customized Xilinx ALVEO™ FPGA board. Moreover, the evaluation of some application layer data relied on NP simulators, which can achieve cycle-accurate performance evaluation.

\paragraph{Workload}

For low-level service mesh testing, the experiment focused on testing indicators including the system's throughput capacity, transmission latency, and transmission stability. Factors that affect the performance of these systems include packet size, burst traffic size, number of connections, and the number of CPU processing cores. We used iperf as a throughput testing tool and sockperf as a delay testing tool. Iperf can simulate different sizes of burst traffic as the test load, while sockperf can simulate the sending and receiving of packets of different sizes and perform delay statistics. For the stability test of transmission, we used iperf sending UDP packets to measure delay jitter.

In our experiment, the application layer service communication mainly focused on the HTTP protocol. The HTTP protocol generally focuses on the ability to respond to requests, including response speed, average response latency, and long tail latency. For this experiment, Fortio  was used as the payload generator. Fortio can send requests at different rates while supporting request sending under different concurrent conditions. Additionally, this experiment also used a unilateral system testing method to test the performance of service and proxy communication optimization in section \ref{lab:communication}. The system was deployed with a system to be tested on one end and a load transceiver program on the other end, using a load generator called wrk. The above load generation and service deployment were both located on the host side, and the DPU was only responsible for processing the transmitted data.

\paragraph{Benchmark systems}

We selected three types of benchmark systems for testing. "envoy": This is the baseline system and deploys service mesh by traditional means \cite{25.istio}. "sockmap": This system uses eBPF to optimize the communication within services and service mesh pods \cite{26.cilium}. "TOE": This system uses existing hardware optimization techniques such as SR-IOV, TOE and OVS offloading to improve communication within the service mesh. We reproduced the three benchmark systems in our experiment and compared them with FlatProxy proposed in this paper.

\subsection{FlatProxy performance}
In this section, we evaluate the effectiveness of our design by comparing the throughput, latency, stability, and CPU utilization of end-to-end communication. 

\begin{figure}[t]
  \centering
  \includegraphics[width=140pt]{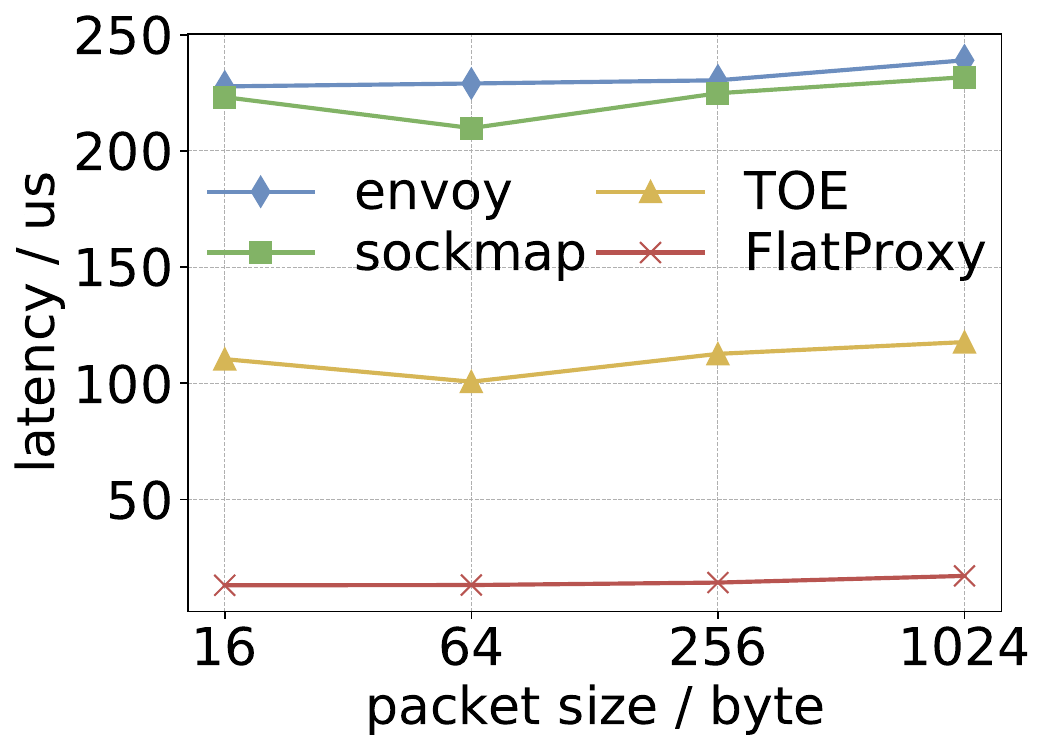}
  \caption{CPU utilization over varying traffic size}
  \label{fig:latency_l3}
\end{figure}

\begin{table}
\begin{center}

\caption{The latency breakdown of data transmission between NIC and service in the L4 \cite{46.AccelTCP}}
% \centering
\begin{tabular}{| l | c | c | l |}
\hline
   envoy & 22us  &  7.6us & FlatProxy  \\
\hline
  vSwitch                     & 9\%  & 26\% & OVS \\
\hline
  TCP/IP protocol             & 27\% & 1\%  & TOE\\
\hline
  \textbf{proxy operation:}   &      &      & \\
\hline
  TCP $\rightarrow$ proxy     & 20\% & --   & TOE $\rightarrow$ MAL\\
\hline
  data processing             & 7\%  & 66\% & match-action logic(MAL)\\
\hline
  proxy $\rightarrow$ TCP     & 10\% & --   & MAL $\rightarrow$ VQ \\
\hline
  loopback     & 27\% & 7\%   & VQ $\rightarrow$ service\\
  
\hline
\end{tabular}
% }
\end{center}
\end{table}

\subsubsection{Lower latency}

Figure \ref{fig:latency_l3} illustrates the achieved latency of different proxy implementations. 
"Sockmap" has only a minor impact on latency. "TOE" reduces the latency from 240us to 120us, thanks to hardware offloading that enhances data processing performance and reduces data copying between container, host, and NIC. However, "TOE" cannot eliminate the latency penalty, which can reach up to 100us, induced by envoy. The figure shows that "FlatProxy," an offloading proxy, decreases the latency by 90\% compared to "envoy." This is because "FlatProxy" only retains the necessary processing shown in Table \ref{tab:l3-breakdown}, which reduces a portion of network processing and software overhead between the network and service. For example, FlatProxy eliminates data copying between TCP/IP protocol and proxy (TOE$\leftrightarrow$proxy), and the loopback communication is replaced by socket direct (VQ$\rightarrow$service). Additionally, the hardware accelerator improves data processing speed. 

\begin{figure}[t]
\centering     %%% not \center
% \subfloat{ 
%         \includegraphics[width=120pt]{figure/evaluation/latency_packet.pdf}
%         \label{fig:latency:a}}
\subfloat{ 
        \includegraphics[width=120pt]{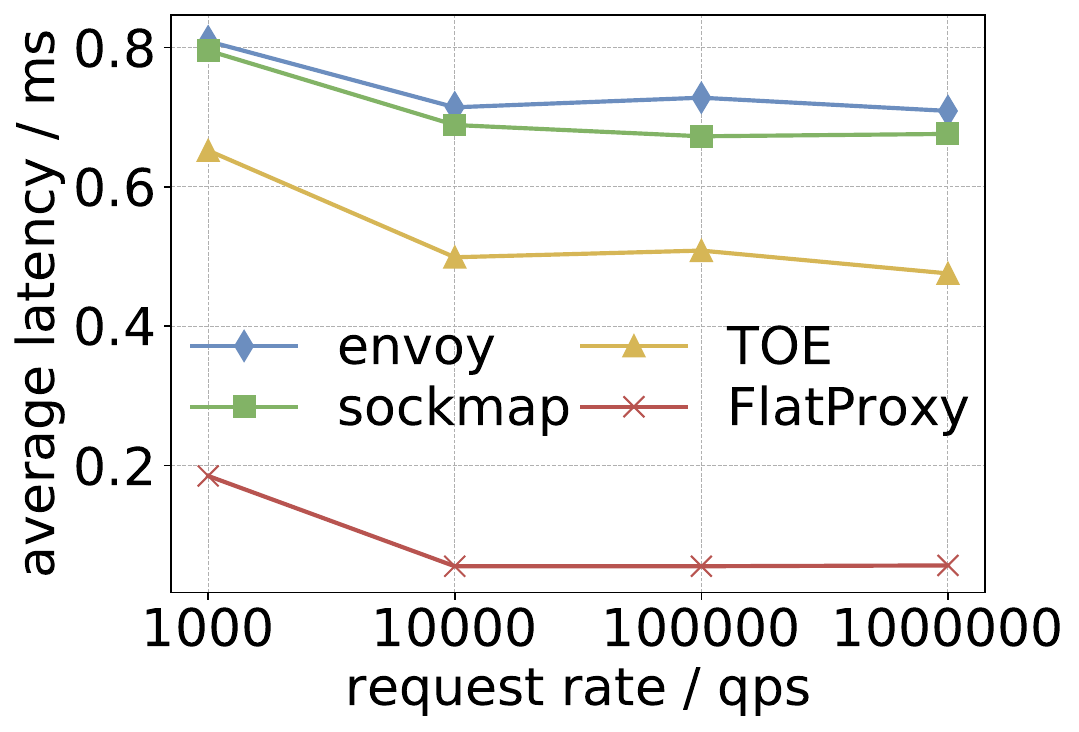}
        \label{fig:latency:b}}
\subfloat{ 
        \includegraphics[width=120pt]{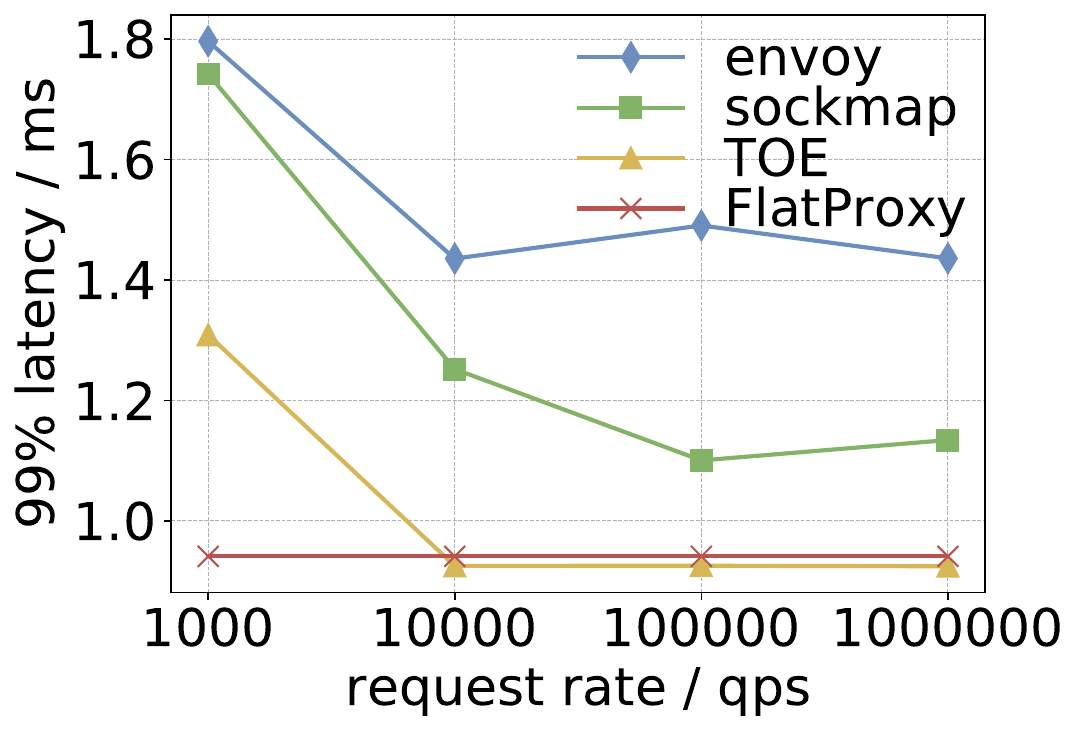}
        \label{fig:latency:c}}        
\caption{Average respond latency and long-tail latency over varying qps rate}
\label{fig:latency}
\end{figure}

\begin{table}
\begin{center}
\caption{The latency breakdown of data transmission between NIC and service in the L7}
\label{tab:l3-breakdown}

% \resizebox{\textwidth}{110pt}{
% \begin{tabular}{p{2.5cm}<{} p{1cm}<{}p{1cm}<{} p{2.5cm}<{}}
\begin{tabular}{ | l | c | c | l | }

\hline
   envoy & 62.5us  &  17.6us & FlatProxy  \\
\hline
  vSwitch                     & 3\%  & 11\% & OVS \\
\hline
  TCP/IP protocol             & 8\% & 1\%  & TOE\\
\hline
% \multicolumn{4}{l}{\textbf{proxy operations:}} \\
\textbf{proxy operations:} & & &\\
\hline
  connection, statistical     & 26\%  & --  & -- \\
\hline
  D-T                         & 16\%  & --  & data transmission \\
\hline
  D-I, D-P                    & 12\%  & 28\% & http parser\\
\hline
  D-CK, D-OP, D-M             & 22\%  & 29\% & match-action logic\\
\hline
  D-DP                        & 6\%   & 28\% & http deparser\\
\hline
  loopback     & 7\% & 3\%   & VQ -> service\\
\hline

\end{tabular}
\end{center}
\end{table}

Figure \ref{fig:latency} presents the average latency and long-tail latency (99\%) of different service mesh optimizations. "FlatProxy" has the lowest average latency, which is for the same reason as the evaluation of TCP communication latency. As shown in Table \ref{tab:l3-breakdown}, FlatProxy eliminates redundant communication paths like loopback and software overheads like data initialization (D-I). Data transmission on the hardware is negligible, which is much faster than host data copying (D-T). Moreover, the data processing, such as HTTP parse/deparse and match-action logic of FlatProxy, is faster than envoy. Additionally, statistical function and connection management functions implemented by the service mesh can decrease performance due to the sequential execution of different operations in the same thread. However, these functions can be implemented in parallel and have little influence on FlatProxy data processing latency. We also found that the average latency is high at the beginning of the request because the proxy needs to initialize some settings at the inception of the request.

Furthermore, we evaluated the long-tail latency (99\%). Figure \ref{fig:latency} shows that "TOE" and "FlatProxy" have the lowest long-tail latency, meaning that hardware offloading can significantly improve the forwarding capability and reduce queue time. The figure also demonstrates that "FlatProxy" can effectively decrease long-tail latency when the request rate varies. The reason why "TOE" has the same latency as "FlatProxy" when the request rate is more than 10000qps may be due to the application's RX/TX queue congestion, caused by the limitation of application response rate. Long-tail latency optimization is not our primary focus, so we did not thoroughly analyze this phenomenon.

\subsubsection{Higher throughput}

\begin{figure*}
\centering     %%% not \center
\subfloat[Throughput over varying traffic]{ 
        \includegraphics[width=120pt]{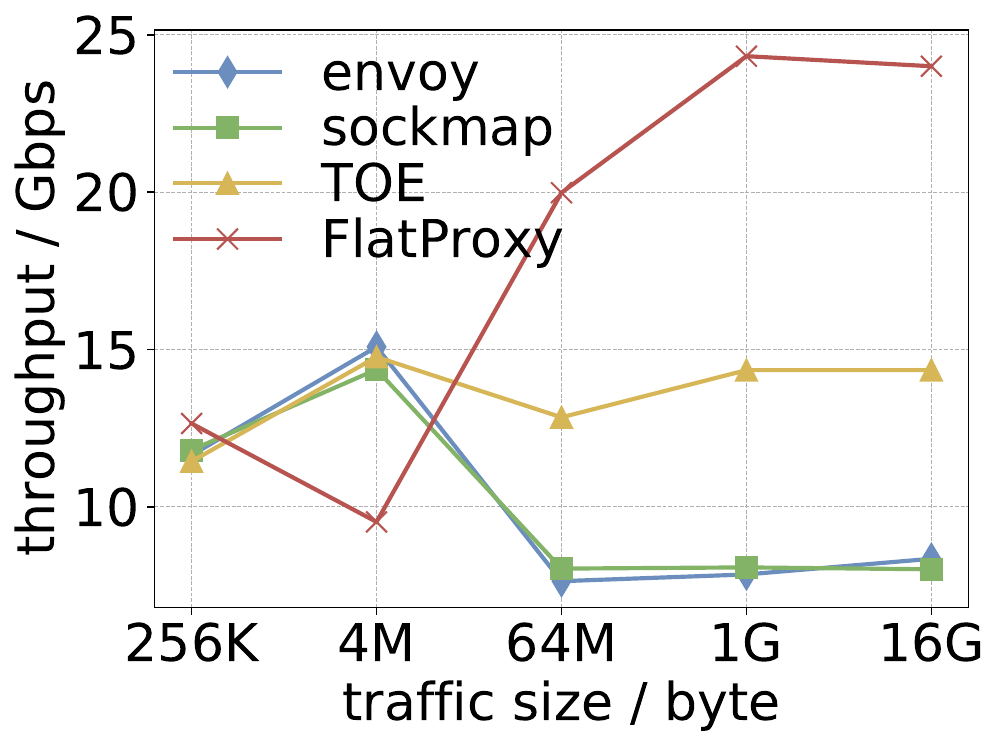}
        \label{fig:throughput:a}}
\subfloat[Throughput over varying connection]{ 
        \includegraphics[width=120pt]{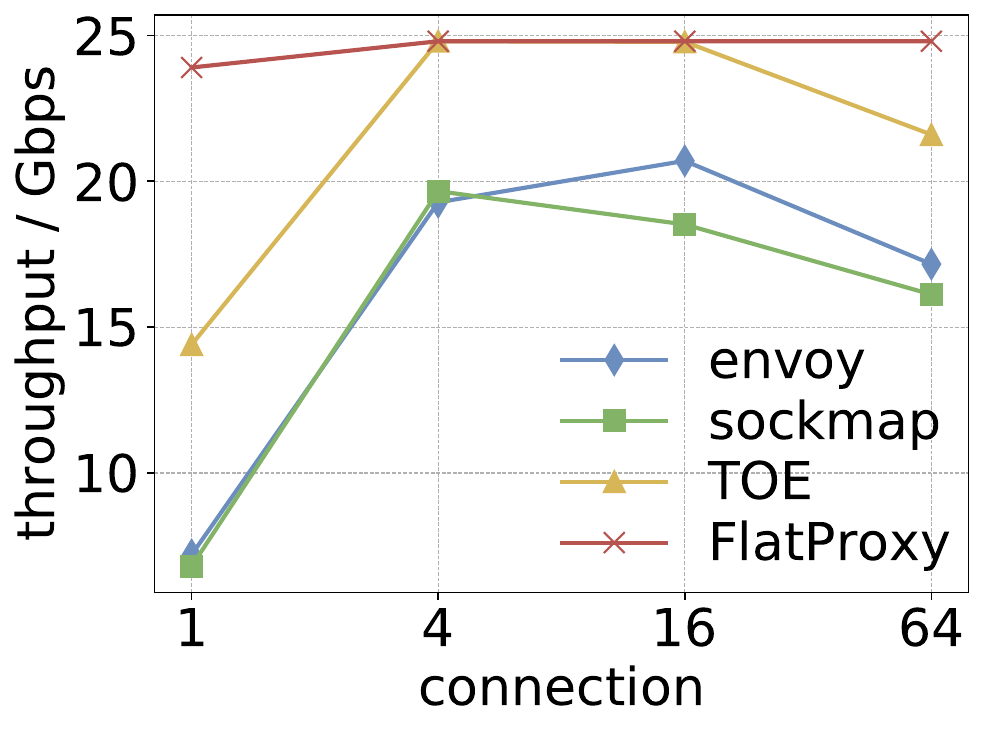}
        \label{fig:throughput:b}}
\subfloat[Throughput over varying core]{ 
        \includegraphics[width=120pt]{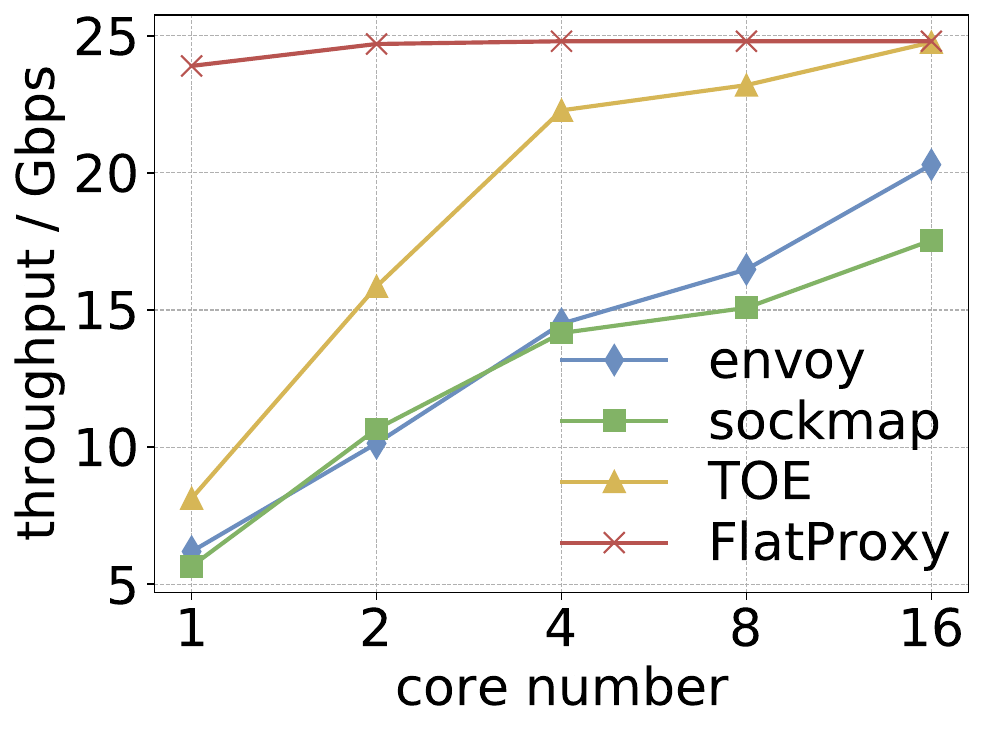}
        \label{fig:throughput:c}} 

\subfloat[Respond rate over varying request rate]{ 
        \includegraphics[width=120pt]{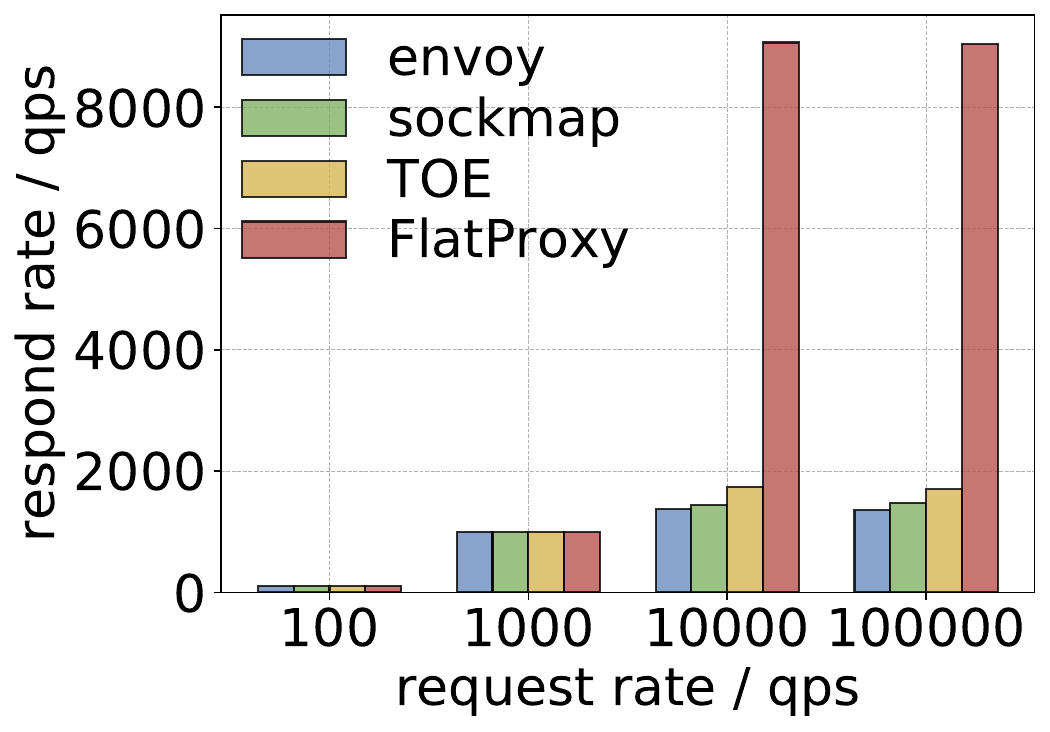}
        \label{fig:qps:a}}
\subfloat[Respond rate over varying connection]{ 
        \includegraphics[width=120pt]{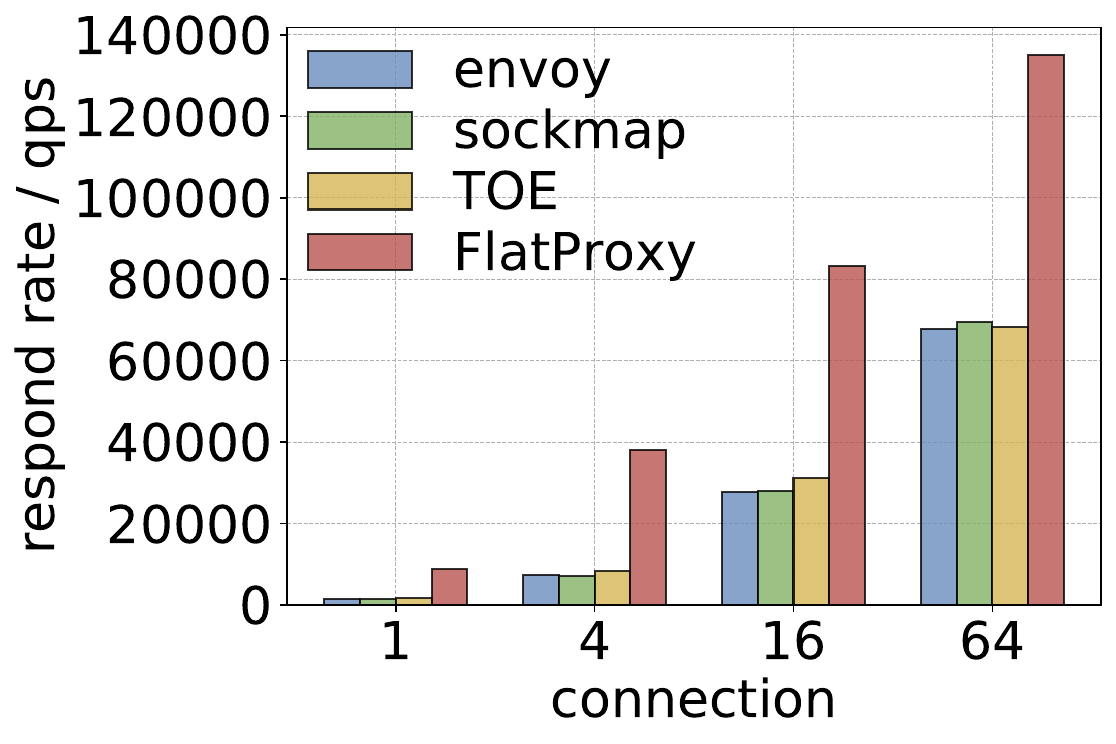}
        \label{fig:qps:b}}
\subfloat[Respond rate over varying core]{ 
        \includegraphics[width=120pt]{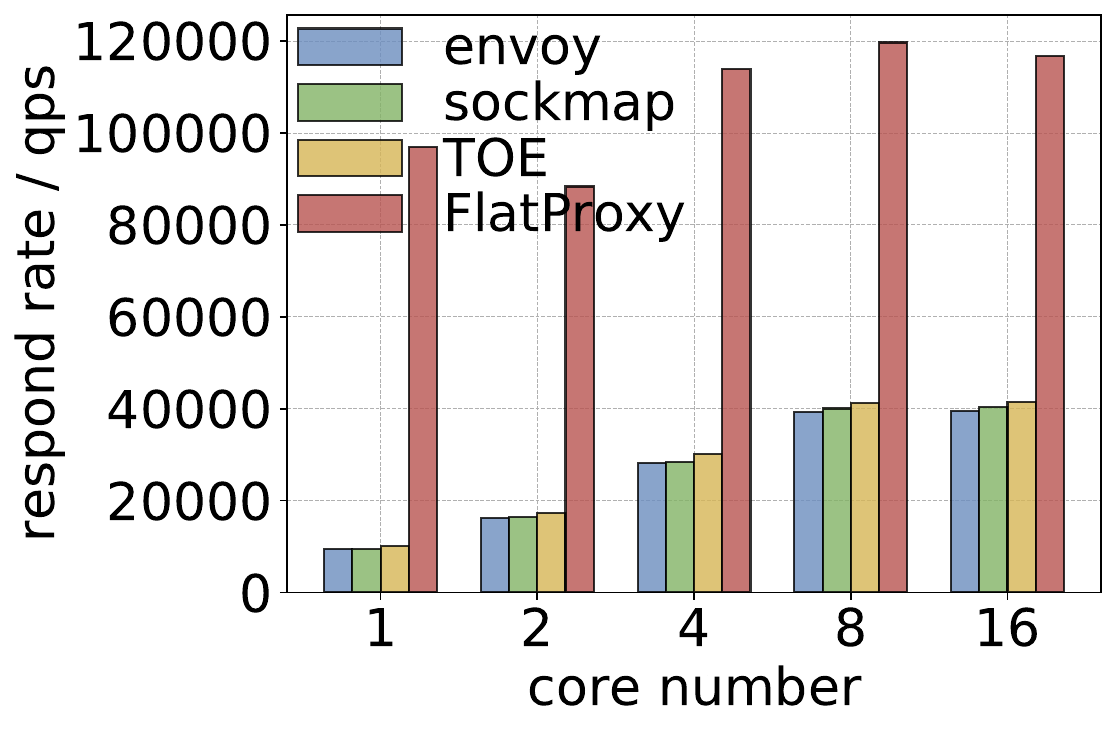}
        \label{fig:qps:c}}
\caption{Average respond latency and long-tail latency over varying qps rate}
\label{fig:throughput}
\end{figure*}

Figure \ref{fig:throughput:a} shows the achieved throughput over varying traffic sizes. "Sockmap" and "envoy" have the same throughput, which indicates that communication within the Pod is not the bottleneck. "TOE" offloading virtual switch and TCP/IP stateless operations increases throughput to 14Gbps, and "FlatProxy" has the highest throughput near line speed (25Gbps), which is 4x higher than "envoy". The high-performance data processing by hardware offloading and hierarchy processing architecture implemented in FlatProxy allows it to save more CPU resources that are used to process data. However, in this evaluation, we observed an abnormal phenomenon where "Flatproxy" has lower throughput when the traffic size is smaller than 4 MB. According to our analysis, this may be due to the statistical method (traffic size/RTT) used by the tool, and the proxy can respond faster to the tool when the traffic is low.

Figure \ref{fig:throughput:b} shows throughput over varying TCP connections. "Sockmap" has similar throughput to "envoy", but its throughput is less than "envoy" when TCP connections are more than four due to the absence of the capability of connection management, as it bypasses the protocol stack between the proxy and service. "TOE" reaches peak throughput when four connections are built, but throughput decreases with increasing connections, especially when there are more than 16 connections. In contrast, "FlatProxy" always maintains the highest throughput and is barely affected by the connections. This phenomenon demonstrates the effectiveness of the hardware multi-thread architecture implemented by FlatProxy, which not only implements dynamic function chains, but also reduces the performance interference between different connections.

Figure \ref{fig:throughput:c} shows throughput over varying CPU cores. "Sockmap" has the lowest throughput due to its bypassing the protocol stack, which lacks the capability of connection management. "FlatProxy" has the highest throughput, which peaks when the testing tool is assigned 2 CPU cores. Theoretically, the RMT pipeline architecture can implement even higher throughput\cite{73.rmt}.

Figures \ref{fig:qps:a}, \ref{fig:qps:b}, and \ref{fig:qps:c} demonstrate the response rate over varying request rate, connection, and CPU core. "FlatProxy" always has the highest response rate in all evaluations, and the performance gap between "FlatProxy" and "envoy" is up to 8 times. In fact, "FlatProxy" can support even higher request processing. A single NP core can complete the request processing in approximately 2000 clocks, and assuming the execution frequency reaches 1 GHz, the prototype platform (with 8 NP cores) can support 4M qps on paper. Meanwhile, these figures also show that the response rate is similar between "envoy", "sockmap", and "TOE", which indicates that a service mesh based on CPU is the bottleneck and that other optimizations have only minor impacts.

\subsubsection{Stability of Comunication}

\begin{figure}
\centering     %%% not \center
\subfloat[Relative latency jitter]{ 
        \includegraphics[width=120pt]{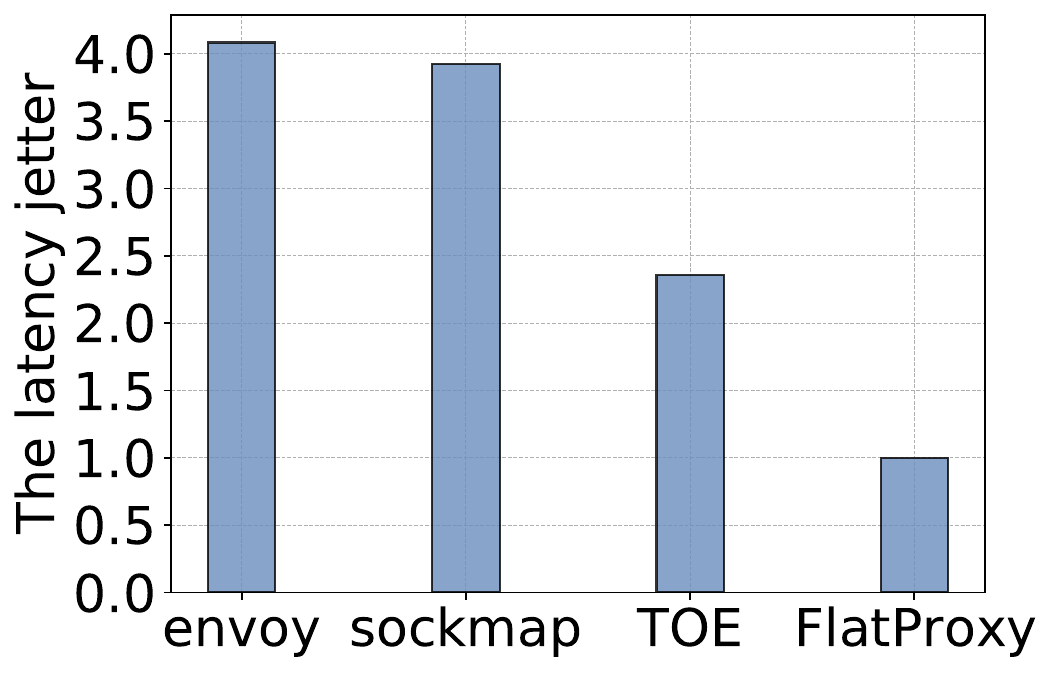}
        \label{fig:stability:a}}
\subfloat[Packet loss]{ 
        \includegraphics[width=120pt]{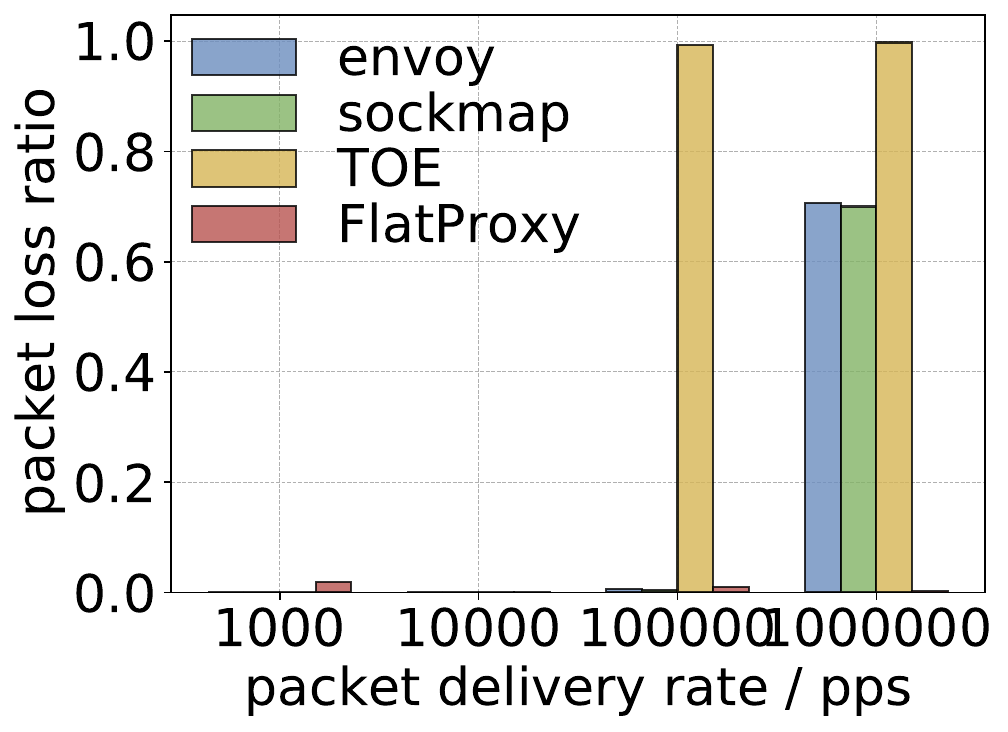}
        \label{fig:stability:b}}
\caption{Stability of communication}
\label{fig:stability}
\end{figure}

\begin{figure*}
\centering     %%% not \center
\subfloat[Percentile latency]{ 
        \includegraphics[width=120pt]{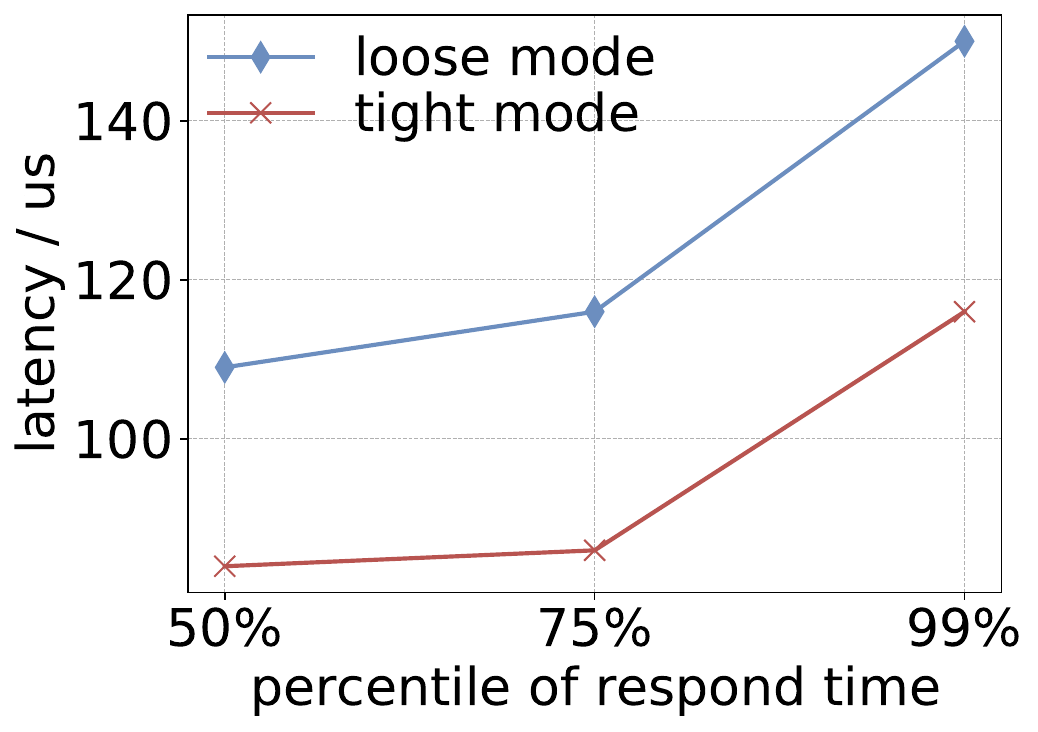}
        \label{fig:communication:a}}
\subfloat[Latency over varying connection]{ 
        \includegraphics[width=120pt]{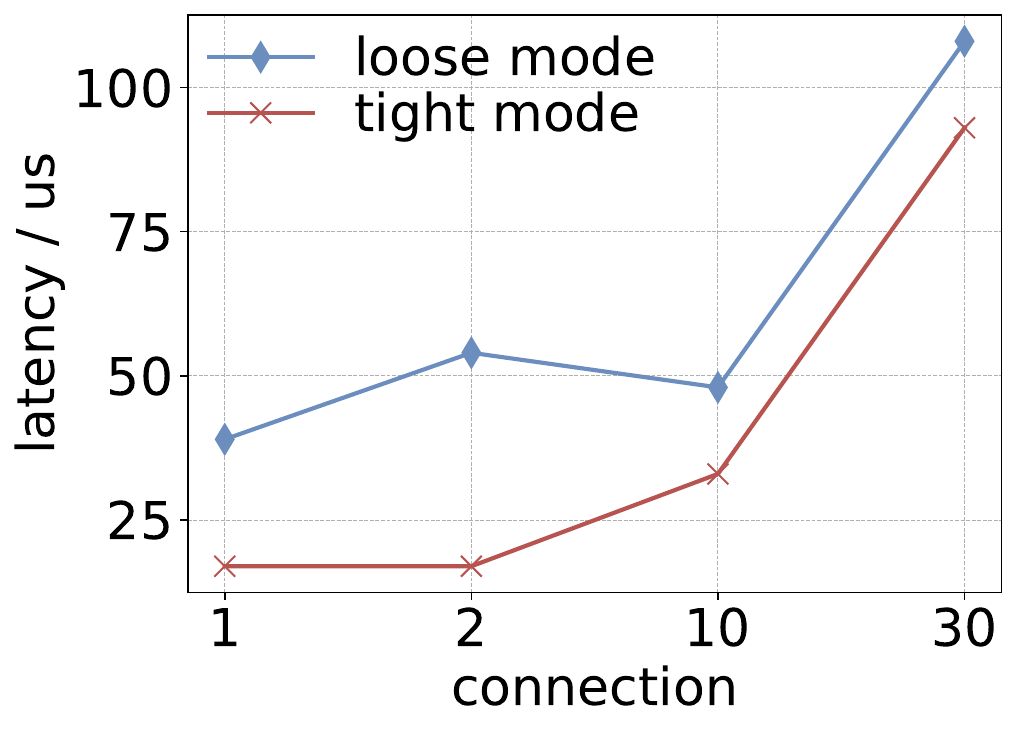}
        \label{fig:communication:b}}
\subfloat[Respond rate]{ 
        \includegraphics[width=120pt]{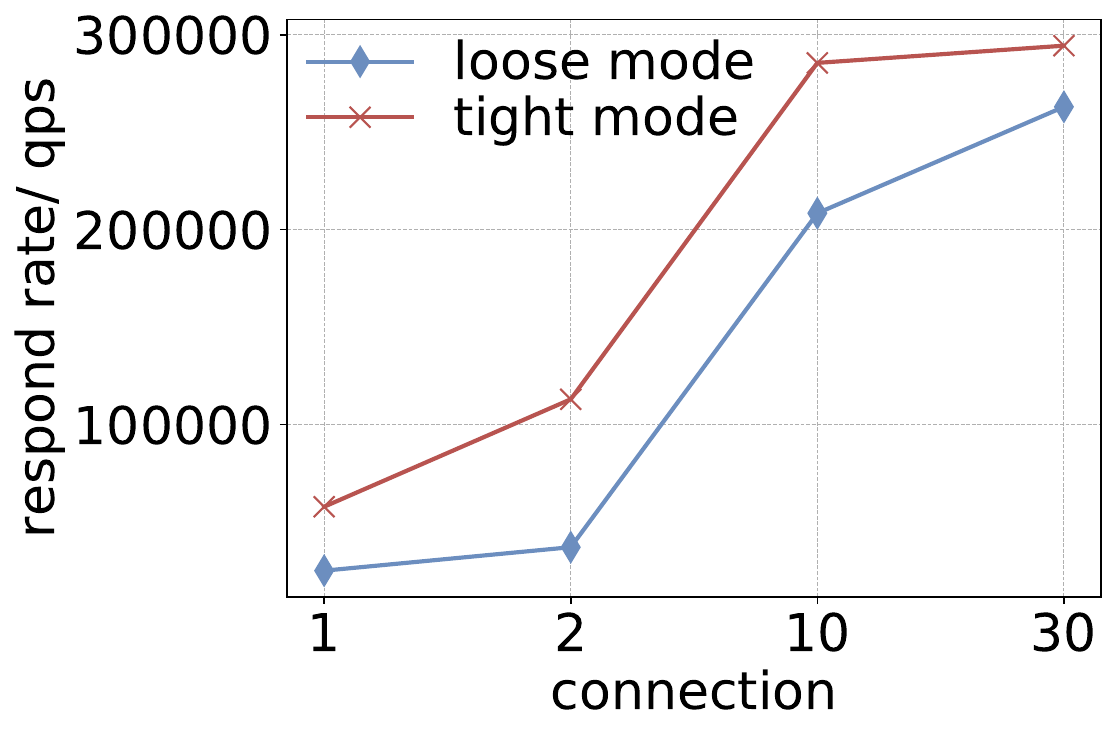}
        \label{fig:communication:c}}        
\caption{The performance of Communication optimization}
\label{fig:communication}
\end{figure*}

The stability of communication is crucial for ensuring the quality of service (QoS). Figure \ref{fig:stability} demonstrates the stability of communication by measuring latency jitter and packet loss. "FlatProxy" exhibits the lowest latency jitter, and Figure \ref{fig:stability:a} illustrates that as the number of software operations decreases, the jitter also becomes smaller. In addition, the evaluation of packet loss indicates that "FlatProxy" also ensures more stable communication. When the request speed exceeds 100,000 qps, "FlatProxy" only loses a few packets compared to baseline solutions. "TOE" has higher packet loss than other optimizations, which indicates that the service mesh based on CPU has weak forwarding capability and is unable to process high-speed packets forwarded by the hardware.

\subsubsection{CPU Utilization}
\begin{figure}[t]
  \centering
  \includegraphics[width=140pt]{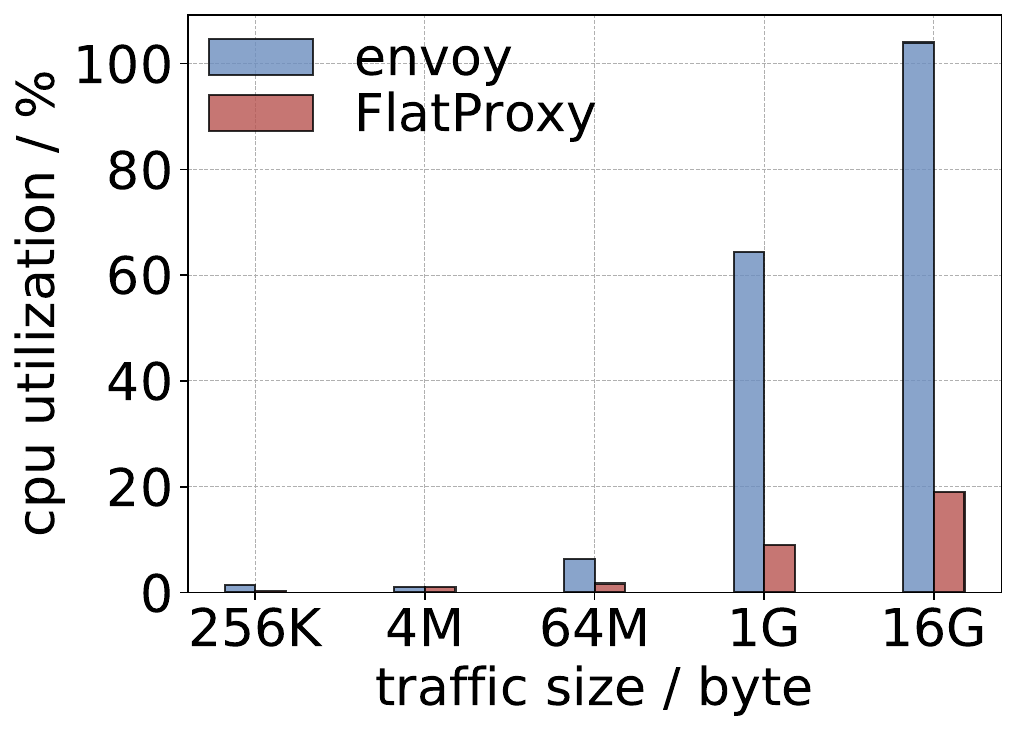}
  \caption{CPU utilization over varying traffic size}
  \label{fig:cpu-utilization}
\end{figure}

Providing more CPU resources to tenants' services is the primary goal of cloud providers. Figure \ref{fig:cpu-utilization} illustrates that "FlatProxy" occupies very little CPU resources and leaves more CPU resources available for the service. In contrast, "envoy" takes up more than 100\% CPU (one core) resources, which is five times more than "FlatProxy". This is because "FlatProxy" isolates the service mesh and service, reducing the CPU utilization of the host. "Flatproxy" can save many CPU resources compared to the traditional envoy solution in the same network bandwidth.

\subsection{Communication Performance} \label{lab:communication}
In this subsection, we validate the effectiveness of the communication optimization proposed in section \ref{communication_opt}. Figure \ref{fig:communication} shows the performance comparison of different communication modes. The tight mode is our optimization, which decreases latency by 20 to 30us compared to the loose mode (container kernel bypass). The response rate is inversely proportional to latency, and the tight mode is superior to the loose mode. Figure \ref{fig:communication:b} also indicates that the communication mode is not the key factor in multi-connection performance interference.

\section{Related work}
Our work aims to implement a lossless service mesh, and we explore the hardware acceleration design using the new hardware platform DPU. Some related work for this topic involves software optimization, developing a suitable hardware platform, and using the hardware to implement related functions.

\textbf{Software Optimization}: Recent service mesh optimization work focuses on reducing data copying and network processing. For example, VPP is used to implement the user-mode network protocol and build the high-performance forwarding plane \cite{15.NSM-VPP}. The open-source project cilium uses eBPF to develop a kernel service mesh function and eliminate redundant communication paths \cite{26.cilium}. Tetrate combines a sidecar proxy and an API gateway to build a unified central management plane \cite{40.Tetrate}. Additionally, different cloud vendors implement customized proxies to improve performance, such as SOFAMesh and Mesher, which optimize routing, scheduling, and customized protocols \cite{41.alibaba, 42.huawei}.

\textbf{Hardware Evolution}: Hardware acceleration is a matter of great concern in the cloud data center. Since Azure proposed AccelNet \cite{45.accelNet}, SmartNIC has made significant progress in academia and industry \cite{44.nitro,51.panic,53.Trio}, with more outstanding data processing and programmable capabilities. The subsequent DPUs provide more robust network, storage, and computing support \cite{18.intel, 19.fungible, 20.nvidia, 21.pensando, 29.yusur}. Google has published the first cloud-native processor, TAU T2A, based on ARM architecture \cite{54.t2a}, which shows a chance for accelerating infrastructure software like service mesh.

\textbf{Service Mesh Offloading}: Currently, Red Hat implements OVN offloading on BF2, which focuses on accelerating functions below the application layer \cite{55.OVN}. Some researches \cite{48.dagger,50.NanoTransport, 52.transport} explore low-latency network protocol implementation on SmartNIC.

\section{Conclusion}
In this paper, we explore a new deployment method for cloud-native applications by proposing a data-centric service mesh. Based on this concept, we design FlatProxy, a new service mesh architecture based on DPU. FlatProxy simplifies communication paths and accelerates data processing through a hardware accelerator. Our evaluation shows that FlatProxy can effectively alleviate performance loss in service communication when introducing service mesh. Additionally, this exploration is also a valuable attempt at the new hardware platform DPU. We believe that a similar device will have more potential in developing infrastructure in the cloud data center.

\section*{Acknowledgment}
This work is supported in part by National Natural Science Foundation of China (NSFC) under grant No.(62002340, 62090020, 61872336), Youth Innovation Promotion Association CAS under grant No.Y201923, and the Strategic Priority Research Program of the Chinese Academy of Sciences under grant No.XDB44030100.  Part of this work is supported by the internship program of YUSUR Technology Co., Ltd. The corresponding author is Guihai Yan (yan@ict.ac.cn).


\vspace{-2mm}

% \bibliographystyle{IEEEtran}

% \end{thebibliography}

\end{document}